\pdfoutput=1
\documentclass[11pt]{article}

\usepackage[preprint]{acl}
\usepackage{times}
\usepackage{latexsym}
\usepackage{cleveref} % Make sure cleveref is loaded
\crefname{figure}{Fig.}{Figs.}  % Change "fig." to "Fig."
\crefname{table}{Table}{Tables}  % Change "table" to "Table"

\usepackage[T1]{fontenc}
\usepackage{array}
\usepackage{amssymb}
\usepackage{pifont}
\usepackage{booktabs}
\usepackage{xspace}
\usepackage{arydshln}
\usepackage{multirow}
\usepackage{changes}
\usepackage{changepage} % Add in the preamble

\usepackage{xcolor}
\definecolor{DarkGreen}{rgb}{0.0, 0.5, 0.0} 

\usepackage[utf8]{inputenc}

\usepackage{microtype}

\usepackage{inconsolata}
\usepackage{enumitem} % Add this in the preamble

\usepackage{graphicx}
\usepackage{svg}
\renewcommand{\footnotesize}{\scriptsize}

\title{DocReRank: Single-Page Hard Negative Query Generation for Training Multi-Modal RAG Rerankers}

\author{
    Navve Wasserman\textsuperscript{1},  Oliver Heinimann\textsuperscript{1}, 
    Yuval Golbari \textsuperscript{1}, Tal Zimbalist \textsuperscript{1} \\
    \textbf{Eli Schwartz\textsuperscript{2}, Michal Irani \textsuperscript{1}} \\
    \textsuperscript{1}Weizmann Institute of Science  \quad \textsuperscript{2} IBM Research Israel
}

\begin{document}
\maketitle

\begin{abstract}
Rerankers play a critical role in multimodal Retrieval-Augmented Generation (RAG) by refining ranking of an initial set of retrieved documents.
Rerankers are typically trained using hard negative mining, whose goal is to select pages for each query which rank high, but are actually irrelevant.
However, this selection process is typically passive and restricted to what the retriever can find in the available corpus, leading to several inherent limitations. These include: limited diversity, negative examples which are often not hard enough, low controllability, and frequent false negatives which harm training.
Our paper proposes an alternative approach: \emph{Single-Page Hard Negative Query Generation}, which goes the other way around. Instead of retrieving negative pages per query, we generate hard negative queries per page. Using an automated LLM-VLM pipeline, and given a page and its positive query, we create hard negatives by rephrasing the query to be as similar as possible in form and context, yet \emph{not} answerable from the page.
This paradigm enables fine-grained control over the generated queries, resulting in diverse, hard, and targeted negatives. It also supports efficient false negative verification. Our experiments show that rerankers trained with data generated using our approach outperform existing models and significantly improve retrieval performance.
\end{abstract}

\begin{figure*}[th!]
    \centering
    \includegraphics[width=0.9\textwidth]{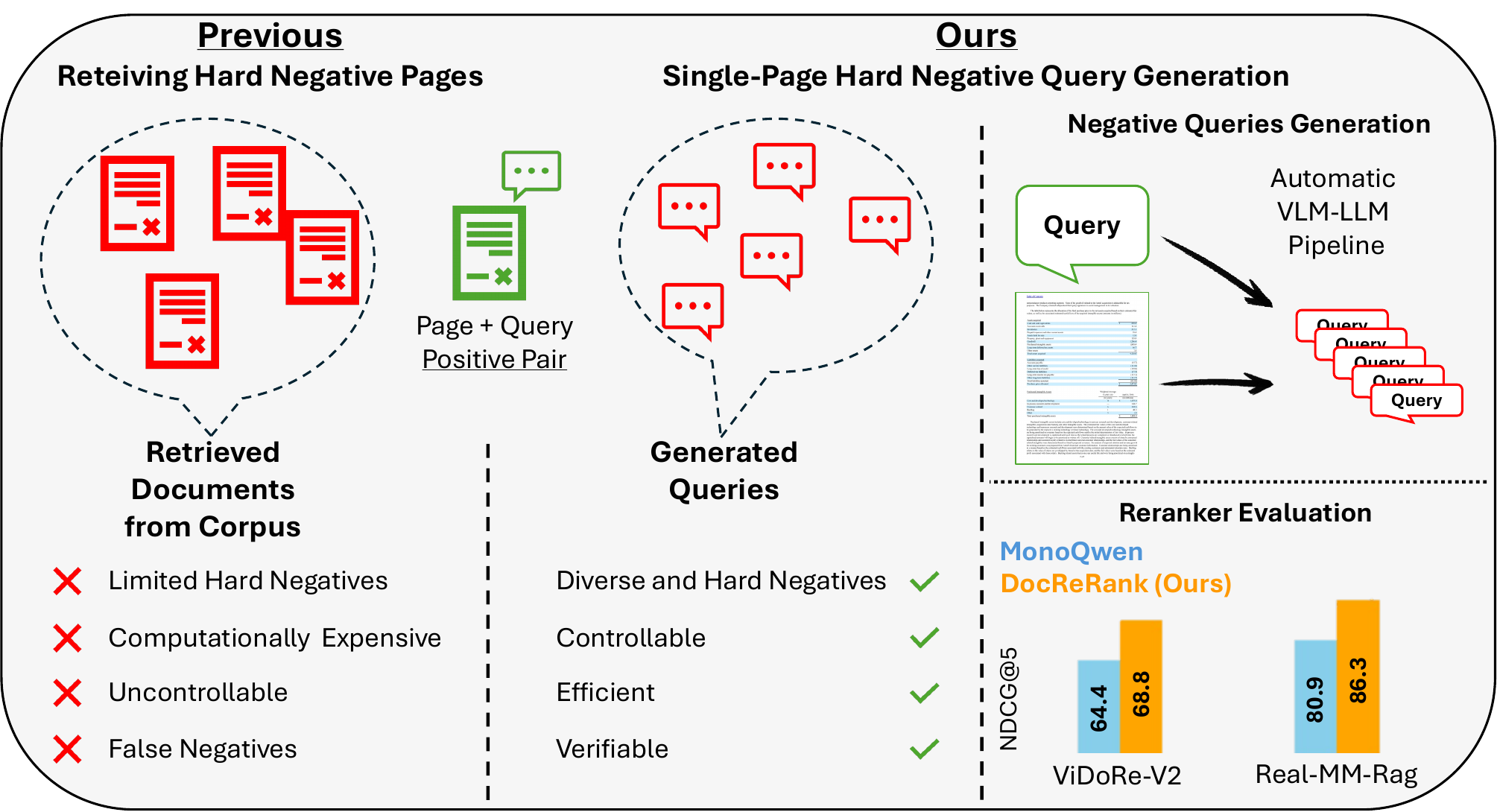} 
    \vspace{-0.21cm}
    \caption{\textbf{Proposed Single-Page Hard Negative Query Generation Approach.} 
    While previous approaches retrieve hard negative pages per query from a document corpus, our method 
    goes the other way around:
    We generate hard negative queries per page using an automated LLM-VLM pipeline. Our reranker, ``DocReRank'' which trains on this kind of 
    data, outperforms models trained with document-based hard negatives.}
    \vspace{-0.43cm}
    \label{fig:Teaser}
\end{figure*}

\vspace{-0.04cm}
\section{Introduction}
\vspace{-0.10cm}

Accurately retrieving relevant documents is fundamental to many natural language processing (NLP) tasks. Retrieval-Augmented Generation (RAG)~\citep{lewis2020retrieval} is a widely adopted framework in which models retrieve external evidence to guide generation. This enables scaling to large document collections while maintaining factual grounding. In real-world applications, \emph{multimodal RAG} extends this framework beyond plain text to include visual and structural elements such as figures, tables, and full-page document images.

While first-stage retrieval models~\citep{ karpukhin2020dense, khattab2020colbert, xiong2020approximate} aim to identify a small set of relevant candidates, their reliance on embedding similarity often limits precision, especially in visually complex settings.
To improve fine-grained relevance, a second-stage \emph{reranker} is commonly used to reorder the top-k documents based on richer query-document interaction.
Reranking has been extensively studied in text-based RAG~\citep{nogueira2019multi, nogueira2020document, sun2023chatgpt, liu2025leveraging}, with one prominent approach adapted to the multimodal setting~\citep{MonoQwen}.

A common training strategy is hard negative mining: for each query, passages or pages labeled as negatives are selected based on their relevance ranking from a retrieval model.
However, this approach faces several key limitations:
(i) \emph{\textbf{Limited hard negatives:}} Negatives are restricted to documents in the corpus, limiting diversity and difficulty;
(ii) \emph{\textbf{Uncontrollable:}} The selection process is passive; only what the retriever pulls out
%surfaces 
can be used, making it hard to target specific model weaknesses (e.g., fine-grained distinctions);
(iii) \emph{\textbf{Computationally expensive:}} The process requires embedding the entire corpus and performing a full retrieval search for each query, making it resource-intensive;
(iv) \emph{\textbf{False negatives:}} Documents incorrectly labeled as irrelevant despite containing the answer are common and can significantly harm training.

We propose an inverse approach: \emph{\textbf{Single-Page Hard Negative Query Generation}}.
Instead of retrieving negative pages per query, we generate hard queries per a given document page.
This approach is inverse not only because it generates rather than retrieves, but also because the negatives are queries instead of documents, avoiding the need to synthesize full pages, which is far more complex and often low quality. Our automated pipeline combines Large Language Models (LLMs) and Vision-Language Models (VLMs) to generate positive queries which are answerable from the page, then rephrases them into hard negatives that are structurally and semantically similar but unanswerable.

This approach addresses the key limitations of document-focused hard negative mining:
(i) \emph{\textbf{Diverse and Hard Negatives:}} By generating queries instead of relying on retrieving documents, we avoid dataset constraints, and can produce diverse, challenging negatives for any page.
(ii) \emph{\textbf{Controllable:}} We explicitly control the type of negative queries generated, allowing us to target specific model weaknesses;
(iii) \emph{\textbf{Efficient:}} Our method eliminates the need to embed and search over large document corpora for each query, significantly reducing the computational cost of hard negative generation;
(iv) \emph{\textbf{Verifiable:}} Since multiple negative queries relate to the same page, VLM-based verification is fast and reliable, reducing false negatives.

We show that training rerankers with data generated by our proposed \emph{Single-Page Hard Negative Query Generation} approach significantly outperforms models trained with document-based hard negatives alone. 
Furthermore, our method can be tailored to address specific model weaknesses. 
For example, we observed that rerankers perform poorly on financial documents and having recurring errors involving fine-grained factual distinctions (e.g., years, numerical values, entity names). Therefore,
we curate a finance-focused dataset using targeted prompts that modify individual attributes during negative query generation. This produces especially challenging negatives that improve reranker robustness in structured, information-dense settings. While finance motivated this effort, such fine-grained variations also appear in other domains, like corporate reports and scientific papers, highlighting the broader applicability of our approach. Training with this dataset yields additional performance gains.

\begin{figure*}[th!]
    \centering
    \includegraphics[width=0.95\textwidth]{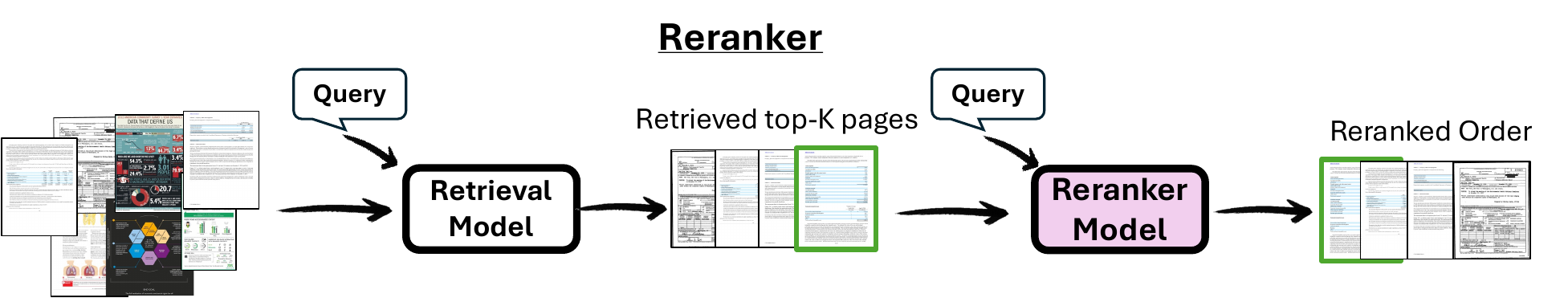} 
    \vspace{-0.1cm}
    \caption{\textbf{Re-ranking Framework.}  Given a query and a document corpus, a retrieval model first retrieves the top-\(K\) relevant pages. A reranker then reorders these \(K\) pages based on the query to improve retrieval quality.}
    \vspace{-0.3cm}
    \label{fig:Reranker}
\end{figure*}

Lastly, we examine the impact of training data quality beyond initial query generation. In the original 
ColPali train-set, many positive queries closely mirror document wording, encouraging shallow keyword matching rather than true semantic understanding. Following the insights of~\citet{wasserman2025real}, we create a rephrased version of the dataset, modifying query phrasing while preserving meaning. 
{Models trained on this}
data show improved performance on standard benchmarks and greater robustness on the  
{rephrased version of the Real-MM-RAG benchmark.}

\vspace{0.23cm}
\noindent
\textbf{Our contributions are as follows:}
\vspace{-0.2cm}
\begin{itemize}[leftmargin=*] 
    \setlength{\itemsep}{3pt}
    \setlength{\parskip}{0pt}
    \setlength{\itemindent}{1pt} 
    \item We propose \emph{Single-Page Hard Negative Query Generation} approach, for creating challenging, controllable, and verifiable hard negatives.
    \item \emph{DocReRank}, a multimodal reranker that outperforms previous models across benchmarks.
    \item \emph{ColHNQue}, a dataset (ColPali Hard Negative Queries) suitable for training  rerankers.
    \item \emph{FinHNQue}, a finance-focused negative queries dataset targeting fine-grained distinctions.
\end{itemize}

%%%%%%%%%%%%%%%%%%%%%%%%%%%%%%%%%%%%%%%%%%%
% Related Work
%%%%%%%%%%%%%%%%%%%%%%%%%%%%%%%%%%%%%%%%%%

\section{Related Work}
In modern information‐retrieval systems, a  first‐stage retriever scans a large corpus to select a handful of candidate documents for a user’s query, and a second‐stage reranker then applies more costly models to reorder those far fewer candidates and boost precision.

\begin{figure*}[th!]
    \centering
    \includegraphics[width=\textwidth]{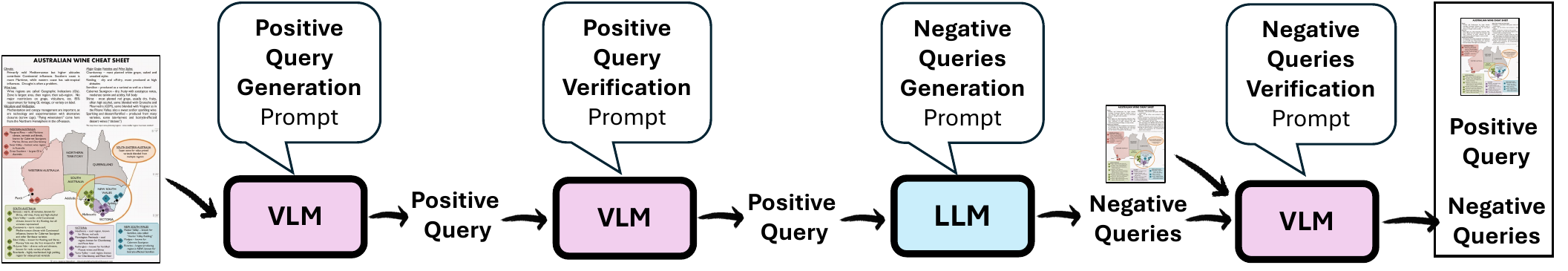} 
    \vspace{-0.5cm}
    \caption{\textbf{Dataset Construction Pipeline.}
    }
    \vspace{-0.38cm}
    \label{fig:Pipeline}
\end{figure*}

\subsection{Retrieval Models}
Early retrieval methods such as TF–IDF \citep{sparck1972statistical} and BM25 \citep{robertson1994okapi} relied on simple lexical matching. These approaches offer extreme efficiency but little semantic understanding. Transformer-based dense retrievers — BERT \citep{devlin2019bert}, T5 \citep{raffel2020exploring}, and DPR \citep{karpukhin2020dense} — map queries and documents into continuous embeddings, dramatically boosting recall at the cost of higher compute. Hybrid retrievers like ColBERT \citep{khattab2020colbert} and ANCE \citep{xiong2020approximate} fuse token-level interactions with vector representations. Yet, text-only retrievers still struggle on richly formatted or visually complex documents.
To bridge that gap, multimodal pipelines are needed. First approaches used captioning-based methods to translate visual elements into natural language \citep{ramos2023retrieval} or contrastive embeddings to align visual and textual features \citep{radford2021learning, zhai2023sigmoid}. 
A more recent line of work leverages the strong capabilities of VLMs to analyze full document images by embedding entire pages, bypassing OCR-based extraction. Methods like VISRAG~\citep{yu2024visrag} and DSE~\citep{ma2024unifying} generate embeddings directly from document images. Similarly, ColPali~\citep{faysse2024colpali} produces multi-vector embeddings for ColBERT-style late interaction retrieval, using PaliGemma~\citep{beyer2024paligemma}, or in its ColQwen {\citep{faysse2024colpali}} variant, Qwen2-VL~\citep{Qwen2VL}. These approaches show clear improvements over earlier methods.

\subsection{Reranking Models}
A Reranker's tasks is to get the top-K candidate documents retrieved in the first stage, and output those documents in a new order, ranked by predicted relevance to the query.
Rerankers can be grouped into three main types: \emph{Pointwise} methods score each document independently given the query (e.g., MonoBERT \citep{nogueira2019multi}, MonoT5 \citep{nogueira2020document}  and CEDR \citep{macavaney2019cedr}). 
\emph{Pairwise} methods compare pairs of documents and predict which one is more relevant, as in DuoT5 \citep{pradeep2021expando}. 
\emph{Listwise} methods optimize over the entire ranked list to capture global ordering subtleties (RankGPT \citep{sun2023chatgpt}, PE-Rank \citep{liu2025leveraging}).

As with retrieval, \emph{multimodal} reranking is needed to handle documents enriched with images, tables, and complex layouts. Specifically, as the best retrieval models operate directly on page images,
these kinds of rerankers are necessary.
While vision–language models can be adapted to judge query–page correspondence, this is far from an optimal solution.
This field is in its early stages with with one prominent model, the MonoQwen~\citep{MonoQwen} reranker, which employs LoRA to fine-tune the Qwen2.5-VL-7B-Instruct VLM using ColPali training data with hard negative mining.

\paragraph{Hard Negative Mining}
Hard negative mining is an important part of effective reranker training, involving the selection of challenging negative examples. 
A trained retrieval model fetches the top-K passages or pages per query, and those not labeled as positive are treated as negatives. 
By identifying difficult negative examples that are semantically similar to the query yet irrelevant, the model learn more discriminative features and improves the training quality.
Early reranker training such as DPR (\citealp{karpukhin2020dense}), BERT passage re‐ranking \citep{nogueira2019passage}, MonoBERT \citep{nogueira2019multi} and ColBERT \citep{khattab2020colbert} 
relied on simple hard negatives, derived from static BM25-mined samples or in-batch examples. Following works adopted dynamic and multi‐retriever mining strategies (e.g., R²anker \citealp{zhou2022towards}) as well as positive-aware hard negative mining \citep[]{moreira2024nv} to further boost performance.
 
However, passage or document-level hard negative mining has several inherent limitations, including limited diversity, false negatives, stale negatives, and little control over the types of negatives retrieved.
To address these limitations, we propose Single-Page Hard Negative Query Generation: a fully automated LLM and VLM pipeline that generates challenging queries for each document page, rather than retrieving hard negative documents for a given question. This paradigm enables fine-grained control, diverse and targeted negatives, and efficient false negative filtering via VLMs, 
since set of negative queries are associated with a single page and can be verified together.
Overall, this produces more challenging and higher-quality training data.

\begin{figure*}[th!]
    \centering
    \includegraphics[width=0.96\textwidth]{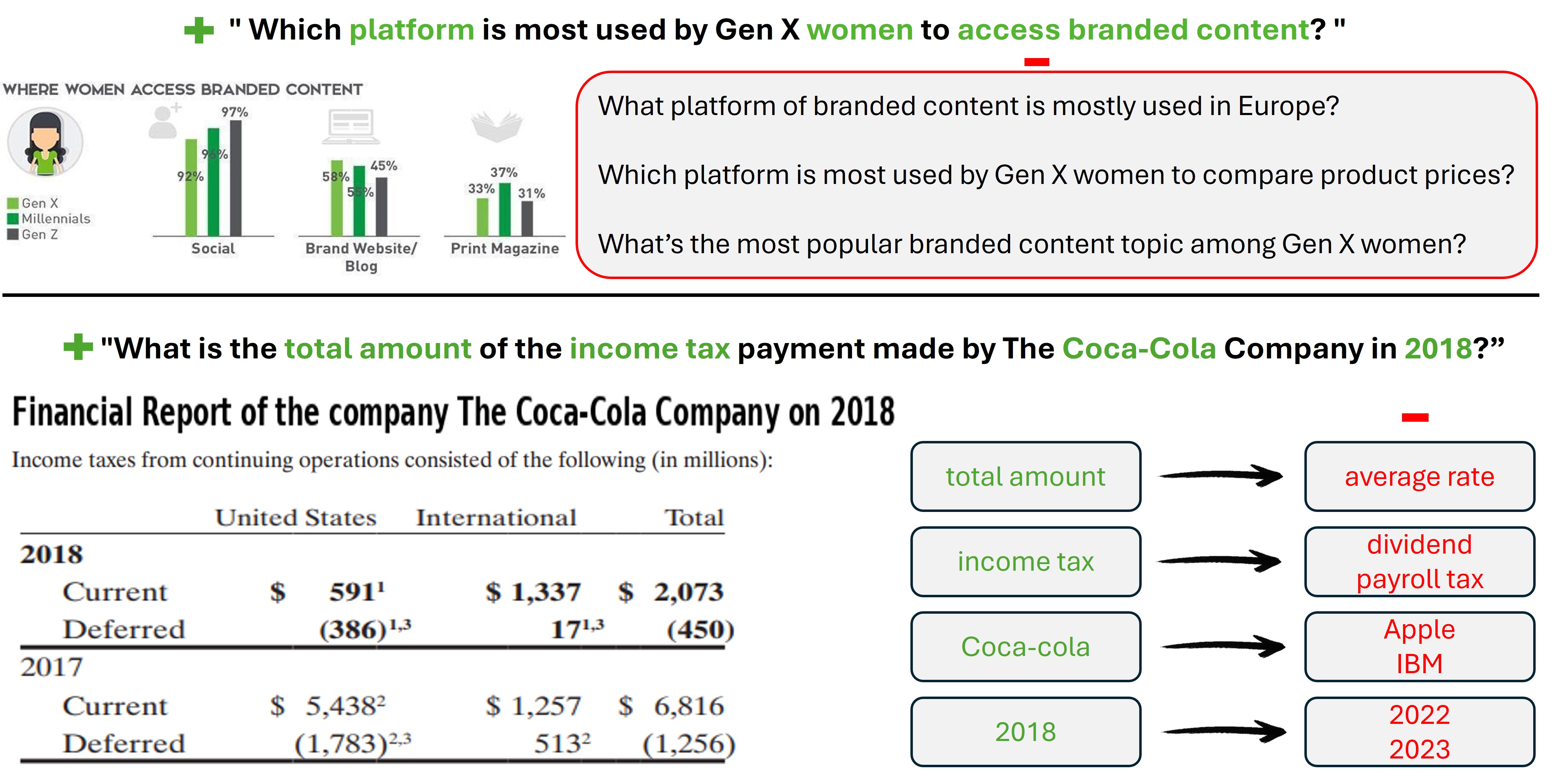} 
    \vspace{-0.3cm}
    \caption{\textbf{Examples of Our Generated Negative Queries.}
    We show examples of a cropped page and its positive query, along with the generated negative queries. Top: hard negatives generated using the general pipeline. Bottom: negatives generated using finance fine-detail prompts, which modify specific properties in the query.}
    \vspace{-0.3cm}
    \label{fig:data_examples}
\end{figure*}

%%%%%%%%%%%%%%%%%%%%%%%%%%%%%%%%%%%%%%%%%%%
% Dataset Generation
%%%%%%%%%%%%%%%%%%%%%%%%%%%%%%%%%%%%%%%%%%
\section{Dataset Generation}
\vspace{-0.12cm}

We propose a new approach for generating document page–query pairs using a dedicated Single-Page Hard Negative Query Generation strategy (see \cref{fig:Pipeline}). Our full pipeline consists of four stages: the first two handle \emph{\textbf{positive query generation and verification}}, while the latter two focus on \emph{\textbf{hard negative query generation and verification}}. If only the hard negative generation is needed (e.g., to extend existing datasets), the process can begin from step 3. Below, we describe the full pipeline and later demonstrate how it can be adapted to model-specific weaknesses.

\subsection{Generation Pipeline}
\vspace{-0.07cm}
Given an image of a document page, the goal is to produce both positive and hard negative queries that relate to the content of that page.

\paragraph{Positive Query Generation} 
We adapt the prompt design from \citet{wasserman2025real} (see \cref{fig_sup:pos_gen_prompt}) and use the Pixtral-12B VLM \citep{agrawal2024pixtral} to generate $N$ candidate positive queries per page. The prompt is designed to encourage RAG-style questions; natural questions that a user might ask without having seen the page itself. It further emphasizes multimodal understanding by focusing on page elements such as figures, tables, and diagrams.
The second stage verifies that each generated query is answerable from the page content.
We use the Qwen2.5-VL-7B-Instruct VLM with a dedicated prompt (see \cref{fig_sup:verif_prompts}) to validate each query. This model is different from the one used for generation, reducing model-specific biases.
After verification, we retain one validated query per page to form a clean set of (page image, positive query) pairs. While multiple positives could be used, we select a single one for simplicity and fair comparison to previous datasets.

\paragraph{Hard Negative Query Generation}
Given a page image and its corresponding positive query, our goal is to generate hard negative queries i.e., queries that are not answerable from the page, but are similar in structure and context to the positive, making them difficult for rerankers to distinguish.
This process is divided into two distinct stages, generation and verification, as we found it significantly more effective to decouple the linguistic task of rephrasing a query from the visual task of grounding it in the page content. Specifically, it is relatively easy for an LLM to generate query variants that are semantically close to the original but seek different information, whereas asking a VLM to handle both rephrasing and verification often led to degraded quality in the resulting negatives.
First, we use the Qwen2.5-7B-Instruct LLM to generate 12 variants of the positive query (see prompt in \cref{fig_sup:neg_gen_prompt}). These are designed to be similar in topic and form but seek different information. This LLM-only step is well-suited for understanding the instruction and generating plausible alternatives. Next, each candidate query is validated using the Qwen2.5-VL-7B-Instruct VLM. We input the document page along with each negative candidate using two slightly different verification prompts (see \cref{fig_sup:verif_prompts}) to improve robustness. Only queries that both prompts classify as unanswerable are kept as valid hard negatives.
This automated pipeline yields high-quality triplets (page image, positive query, hard negatives), which can then be used to train reranker models more effectively.

\subsection{Finance Focused Generation} 
\label{DataGen_Fin}
A key advantage of our proposed negative query generation approach is its adaptability to specific document types. 
Although the commonly used ColPali training set contains a variety of financial documents, the model performance on financial benchmarks remains notably lower. We further observed from error analysis conducted on the Real-MM-RAG benchmark, that models have difficulties in handling fine-grained information distinctions (see \cref{fig_sup:fail_cases_fin_1,fig_sup:fail_cases_fin_2}).

\begin{table*}[ht!]
\footnotesize
\renewcommand{\arraystretch}{1.5} % Adjust row height
\setlength{\dashlinedash}{0.5pt}
\setlength{\dashlinegap}{0.5pt}
\setlength\tabcolsep{2.7pt} % Adjust column spacing
% \hspace{-0.15cm}
\begin{tabular}{@{\extracolsep{\fill}}lccccccccc} 
% \begin{tabular*}{0.51\textwidth}{@{\extracolsep{\fill}}lcccc} 
\toprule
\textbf{Benchmark} & \textbf{Axa} & \textbf{Economics} & \textbf{Restaurant-rse} & \textbf{Restaurant-esg}  & \textbf{Biomedical}& \textbf{Economics-ML}& \textbf{Restaurant-ML}& \textbf{Biomedical-ML} & \textbf{Avg}\\
\midrule
\textit{ColPali}  &55.0	&53.4	&51.5	&55.2	&57.8	&47.6	&52.5	&55.6	&53.6 \\

\hspace{0.05cm} \textit{Qwen-VLM}   & 61.8 {\tiny \textcolor{DarkGreen}{+6.8}}  &	47.9 {\tiny \textcolor{red}{-5.5}}  &	57.7 {\tiny \textcolor{DarkGreen}{+6.2}}  &	60.8 {\tiny \textcolor{DarkGreen}{+5.6}}  &	56.4 {\tiny \textcolor{red}{-1.4}}  &	49.8 {\tiny \textcolor{DarkGreen}{+2.2}}  &	58.0 {\tiny \textcolor{DarkGreen}{+5.5}}  &	56.1 {\tiny \textcolor{DarkGreen}{+0.5}}  &	56.1 {\tiny \textcolor{DarkGreen}{+2.5}}  \\

\hspace{0.05cm} \textit{MonoQwen}   & 68.0 {\tiny \textcolor{DarkGreen}{+13.0}} &	57.8 {\tiny \textcolor{DarkGreen}{+4.4}} &	58.2 {\tiny \textcolor{DarkGreen}{+6.7}} &	68.5 {\tiny \textcolor{DarkGreen}{+13.3}} &	65.9 {\tiny \textcolor{DarkGreen}{+8.1}} &	56.7 {\tiny \textcolor{DarkGreen}{+9.1}} &	59.2 {\tiny \textcolor{DarkGreen}{+6.7}} &	62.6 {\tiny \textcolor{DarkGreen}{7.0}} &	62.1 {\tiny \textcolor{DarkGreen}{+8.5}} \\

\hspace{0.05cm} \textit{\textbf{DocReRank}-Base}   & \underline{\textbf{71.8}} {\tiny \textcolor{DarkGreen}{+16.8}}  &	\textbf{64.7} {\tiny \textcolor{DarkGreen}{+11.3}}  &	\textbf{58.3} {\tiny \textcolor{DarkGreen}{+6.8}}  &		\textbf{68.8} {\tiny \textcolor{DarkGreen}{+13.6}}  &	\underline{\textbf{68.0}} {\tiny \textcolor{DarkGreen}{+10.2}}  &	\textbf{58.3} {\tiny \textcolor{DarkGreen}{+10.7}}  &	\textbf{61.2} {\tiny \textcolor{DarkGreen}{+8.7}}  &	\textbf{62.9} {\tiny \textcolor{DarkGreen}{+7.3}}  &	\textbf{64.2} {\tiny \textcolor{DarkGreen}{+10.7}} \\

\hspace{0.05cm} \textit{\textbf{DocReRank}-Full}   & \textbf{70.7} {\tiny \textcolor{DarkGreen}{+15.7}}  &	\underline{\textbf{67.5}} {\tiny \textcolor{DarkGreen}{+14.1}}  &	\underline{\textbf{63.4}} {\tiny \textcolor{DarkGreen}{+11.9}}  &	\underline{\textbf{71.8}} {\tiny \textcolor{DarkGreen}{+16.6}}  &	\textbf{66.8} {\tiny \textcolor{DarkGreen}{+9.0}}  &	\underline{\textbf{62.8}} {\tiny \textcolor{DarkGreen}{+15.2}}  &	\underline{\textbf{63.3}} {\tiny \textcolor{DarkGreen}{+10.8}}  &	\underline{\textbf{63.7}} {\tiny \textcolor{DarkGreen}{+8.1}}  &	\underline{\textbf{66.2}} {\tiny \textcolor{DarkGreen}{+12.7}}  \\

\addlinespace
% \cmidrule(lr){1-5}
\textit{ColQwen}   & 64.4 & 61.3 & 54.4 & 61.2 & 62.2 & 52.6 & 56.0 & 56.8 & 58.6 \\

\hspace{0.05cm} \textit{Qwen-VLM}   & 66.1 {\tiny \textcolor{DarkGreen}{+1.7}}  & 51.9 {\tiny \textcolor{red}{-9.4}}  & 63.6 {\tiny \textcolor{DarkGreen}{+9.2}}  & 64.2 {\tiny \textcolor{DarkGreen}{+3.0}}  & 57.7 {\tiny \textcolor{red}{-4.5}}  & 50.4 {\tiny \textcolor{red}{-2.2}}  & 62.9 {\tiny \textcolor{DarkGreen}{+6.9}}  & 56.3 {\tiny \textcolor{red}{-0.5}} & 59.1 {\tiny \textcolor{DarkGreen}{+0.5}}\\

\hspace{0.05cm} \textit{MonoQwen}   & 71.4 {\tiny \textcolor{DarkGreen}{+7.0}}  & 60.3 {\tiny \textcolor{red}{-1.0}}  & 61.6 {\tiny \textcolor{DarkGreen}{+7.2}}  & \textbf{72.3} {\tiny \textcolor{DarkGreen}{+11.1}}  & 66.7 {\tiny \textcolor{DarkGreen}{+4.5}}  & 58.2 {\tiny \textcolor{DarkGreen}{+5.6}}  & 61.8 {\tiny \textcolor{DarkGreen}{+5.8}}  & 63.1 {\tiny \textcolor{DarkGreen}{+6.3}} & 64.4 {\tiny \textcolor{DarkGreen}{+5.8}}\\

\hspace{0.05cm} \textit{\textbf{DocReRank}-Base}   & \underline{\textbf{77.3}} {\tiny \textcolor{DarkGreen}{+12.9}}  & \textbf{67.5} {\tiny \textcolor{DarkGreen}{+6.2}}  & \textbf{63.3} {\tiny \textcolor{DarkGreen}{+8.9}}  & 71.7 {\tiny \textcolor{DarkGreen}{+10.5}}  & \textbf{68.5} {\tiny \textcolor{DarkGreen}{+6.3}}  & \textbf{60.7} {\tiny \textcolor{DarkGreen}{+8.1}}  & \textbf{65.2} {\tiny \textcolor{DarkGreen}{+9.2}}  & \textbf{63.8} {\tiny \textcolor{DarkGreen}{+7.0}} & \textbf{67.2} {\tiny \textcolor{DarkGreen}{+8.6}}\\
     
\hspace{0.05cm} \textit{\textbf{DocReRank}-Full}   & \textbf{75.3} {\tiny \textcolor{DarkGreen}{+10.9}}  & \underline{\textbf{68.7}} {\tiny \textcolor{DarkGreen}{+7.4}}  & \underline{\textbf{66.6}} {\tiny \textcolor{DarkGreen}{+12.2}}  & \underline{\textbf{75.8}} {\tiny \textcolor{DarkGreen}{+14.6}}  & \underline{\textbf{68.2}} {\tiny \textcolor{DarkGreen}{+6.0}}  & \underline{\textbf{64.7}} {\tiny \textcolor{DarkGreen}{+12.1}}  & \underline{\textbf{66.0}} {\tiny \textcolor{DarkGreen}{+10.0}}  & \underline{\textbf{64.7}} {\tiny \textcolor{DarkGreen}{+7.9}} & \underline{\textbf{68.8}} {\tiny \textcolor{DarkGreen}{+10.1}}\\

\bottomrule
\end{tabular}
\caption{
\textbf{Model Performance on the ViDoReV2 Benchmark.}  
Retrieval NDCG@5 Results. The first row in each block shows first-step retrieval results using ColPali or ColQwen. The remaining rows correspond to second-step reranking results.
Our model \emph{DocReRank-Base} is trained with a similar configuration to MonoQwen but includes our generated data. \emph{DocReRank-Full} is trained with generated fine-grained details 
and rephrased negative queries.}
\label{Table:vidoreV2}
\vspace{-0.4cm}
\end{table*}

While these issues are most prominent in financial documents, such fine-grained errors (e.g.\ confusing numerical values, time periods, or entity names) can also occur in other document types that include structured or data-rich content (e.g., restaurant annual reports or corporate filings). Therefore, improving robustness to small variations in factual details can benefit a wide range of use cases involving factual or financial information.

To address this, we developed a dedicated set of prompts (see \cref{fig_sup:fin_neg_gen_prompt}) that, given a positive query, instructs the model to generate a variant by modifying exactly one property; such as the year (e.g., 2022 → 2024), company name (e.g., Apple → IBM), numerical value (e.g., price, percentage), financial metric (e.g., revenue, sales, acquisitions), subject metric (e.g., dividends, stocks, options), or business segment (e.g., cloud, software, manufacturing). This produces highly targeted hard negatives that challenge the model's ability to distinguish fine-grained but critical details.

We applied this method to the FinTabNet dataset~\citep{zheng2020global}, which contains annual reports from S\&P 500 companies, generating a training set of 20K pages paired with corresponding positive and domain-specific hard negative queries.

\subsection{Rephrased Dataset}
To improve model semantic understanding and robustness, we introduce a rephrased version of the ColPali training set.
We rephrase 50\% of the positive queries while preserving their meaning using an LLM. This encourages the model to rely on semantic understanding rather than surface-level cues.

%%%%%%%%%%%%%%%%%%%%%%%%%%%%%%%%%%%%%%%%%%%
% DocReRank Training
%%%%%%%%%%%%%%%%%%%%%%%%%%%%%%%%%%%%%%%%%%
\section{DocReRank Training}

Our \emph{DocReRank} reranker is based on the pretrained Vision-Language Model Qwen2-VL-2B-Instruct~\citep{Qwen2VL}, using the same Low-Rank Adaptation (LoRA)~\citep{hu2022lora} configuration as in the ColPali paper~\citep{faysse2024colpali}. LoRA is applied to the transformer layers of the LLM, while the visual encoder is kept frozen.
The reranker is trained on triplets of (query, document page image, label), where the label is 1 if the image contains the answer to the query (positive), and 0 otherwise (negative). We follow the MonoQwen~\citep{MonoQwen} training framework, where the model is prompted to generate the token "True" or "False" given a query and an image. During training, a softmax over the logits of these tokens provides a relevance score, used both as the loss and as the basis for reranking during inference.

\subsection{Training Datasets}

We use three types of training data: (i) standard hard-negative-mined data of document pages,
(ii) data generated by our proposed Single-Page Hard Negative Query Generation approach , and (iii) Rephrasing Variants.

\textbf{Document Page Hard Negative Mining (\textit{Col-HNDoc}):} 
We use a version of the ColPali training dataset, also used in MonoQwen, with hard negatives provided by Nomic-AI~\cite{nomicai_colpali_queries_2025}
(MonoQwen hard negatives mining is not available). Each query is paired with one positive page and three hard negative pages sampled from the top-10 retrieval results. This yields approximately 120k positive pairs and 360k negative pairs, totaling around 480k training examples.

\textbf{Single-Page Hard Negative Query Generation:}
Our generated datasets include two variants: (i) \textit{Col-HNQue} (ColPali Hard Negative Queries)  Based on the same ColPali training set, we keep the original query–positive page pairs and generate three hard negative queries for each page. This dataset is matched in size to \textit{Col-HNDoc}, with the only difference being in the method of generating negatives. (ii) \textit{Fin-HNQue} (Finance Hard Negative Queries): We apply our full query generation pipeline to 20k pages from the FinTabNet dataset~\citep{zheng2020global}, generating one positive query and three hard negatives per page. This results in 80k training examples tailored to the financial domain and to fine-grained
information distinctions (see \cref{DataGen_Fin}).

\textbf{Rephrasing  Variants:}
We further introduce an augmented versions of both \textit{Col-HNDoc} and \textit{Col-HNQue} (\textit{Reph-}) where 50\% of the positive queries are rephrased while preserving their meaning. 

\subsection{Training Procedure}

All models were trained using the Hugging Face Trainer with a learning rate of 1e-4, 100 warm-up steps, and a learning rate decay schedule. Training was conducted on 4 NVIDIA L40S GPUs, with each GPU processing a batch size of 32 examples per step. Each batch consists of 8 positive (image, query) pairs and 24 corresponding negatives.

This batch structure ensures that each positive example is accompanied by its respective negatives within the same batch. For the \textit{ColHNDoc} dataset, a positive consists of a query and its corresponding document page, while the negatives are three other document pages that are hard negatives for the same query. For our proposed datasets, each positive consists of a document page and a positive query, and the negatives are three hard negative queries generated for that page.

We use cross-entropy loss over the softmax probabilities of the "True" and "False" token logits. To address class imbalance between positives and negatives in each batch, we assign a weight ratio of 3:1 in favor of the positive examples. All models are trained for one epoch, with the number of training steps determined by the size of each dataset.

\begin{table}[th!]
\footnotesize
\renewcommand{\arraystretch}{1.5} % Adjust row height
\setlength{\dashlinedash}{0.5pt}
\setlength{\dashlinegap}{0.5pt}
\setlength\tabcolsep{2.2pt} % Adjust column spacing
% \hspace{-0.15cm}
\begin{tabular}{@{\extracolsep{\fill}}lccccc} 
% \begin{tabular*}{0.51\textwidth}{@{\extracolsep{\fill}}lcccc} 
\toprule
\textbf{Benchmark} & \textbf{FinReport} & \textbf{FinSlides} & \textbf{TechReport} & \textbf{TechSlides}  & \textbf{Avg}\\
\midrule
\textit{ColPali} & 52.9 &	62.7&	80.4&	89.4&	71.4\\
		
\hspace{0.05cm} \textit{Qwen-VLM}  & 62.6 {\tiny \textcolor{DarkGreen}{+9.7}}  & 77.0 {\tiny \textcolor{DarkGreen}{+14.3}}  & 73.9 {\tiny \textcolor{red}{-6.5}}  & 78.8 {\tiny \textcolor{red}{-10.6}}  & 73.1 {\tiny \textcolor{DarkGreen}{+1.7}}\\
			
\hspace{0.05cm} \textit{MonoQwen}  & \textbf{73.0} {\tiny \textcolor{DarkGreen}{+20.1}}  & 82.1 {\tiny \textcolor{DarkGreen}{+19.4}}  &  79.4 {\tiny \textcolor{red}{-1.0}}  & 91.9 {\tiny \textcolor{DarkGreen}{+2.5}}  & 81.6 {\tiny \textcolor{DarkGreen}{+10.2}}\\
			
\hspace{0.05cm} \textit{\textbf{DocReRank}-B}  & 71.6 {\tiny \textcolor{DarkGreen}{+18.7}}  & \textbf{86.3} {\tiny \textcolor{DarkGreen}{+23.6}}  & \textbf{89.6} {\tiny \textcolor{DarkGreen}{+9.2}}  & \textbf{94.1} {\tiny \textcolor{DarkGreen}{+4.7}}  & \textbf{85.4} {\tiny \textcolor{DarkGreen}{+14.0}}\\

\hspace{0.05cm} \textit{\textbf{DocReRank}-F}  & \underline{\textbf{73.2}} {\tiny \textcolor{DarkGreen}{+20.3}}  & \underline{\textbf{86.8}} {\tiny \textcolor{DarkGreen}{+24.1}}  & \underline{\textbf{89.5}} {\tiny \textcolor{DarkGreen}{+9.1}}  & \underline{\textbf{94.4}} {\tiny \textcolor{DarkGreen}{+5.0}}  & \underline{\textbf{86.0}} {\tiny \textcolor{DarkGreen}{+14.6}}\\

\addlinespace
% \cmidrule(lr){1-5}
\textit{ColQwen}   & 60.8 & 58.7 & 84.4 & 91.2 & 73.8 \\

\hspace{0.05cm} \textit{Qwen-VLM}  & 65.5 {\tiny \textcolor{DarkGreen}{+4.7}}  & 72.1 {\tiny \textcolor{DarkGreen}{+13.4}}  & 73.4 {\tiny \textcolor{red}{-11.0}}  & 79.1 {\tiny \textcolor{red}{-12.1}}  & 72.5 {\tiny \textcolor{red}{-1.3}}\\
	
\hspace{0.05cm} \textit{MonoQwen}  & 75.6 {\tiny \textcolor{DarkGreen}{+14.8}}  & 76.2 {\tiny \textcolor{DarkGreen}{+17.5}}  & 79.4 {\tiny \textcolor{red}{-5.0}}  & 92.3 {\tiny \textcolor{DarkGreen}{+1.1}}  & 80.9 {\tiny \textcolor{DarkGreen}{+7.1}}\\
		
\hspace{0.05cm} \textit{\textbf{DocReRank}-B}  & \textbf{76.8}{\tiny \textcolor{DarkGreen}{+16.0}}  & \textbf{80.7} {\tiny \textcolor{DarkGreen}{+22.0}}  & \textbf{90.0} {\tiny \textcolor{DarkGreen}{+5.6}}  & \textbf{94.7} {\tiny \textcolor{DarkGreen}{+3.5}}  & \textbf{85.6} {\tiny \textcolor{DarkGreen}{+11.8}}\\

\hspace{0.05cm} \textit{\textbf{DocReRank}-F}  & \underline{\textbf{79.0}} {\tiny \textcolor{DarkGreen}{+18.2	}}  & \underline{\textbf{80.9}} {\tiny \textcolor{DarkGreen}{+22.2	}}  & \underline{\textbf{90.4}} {\tiny \textcolor{DarkGreen}{+6.0	}}  & \underline{\textbf{94.8}} {\tiny \textcolor{DarkGreen}{+3.6	}}  & \underline{\textbf{86.3}} {\tiny \textcolor{DarkGreen}{+12.5}}\\

\bottomrule
\end{tabular}
\vspace{-0.15cm}
\caption{
\textbf{Model Performance on the Real-MM-RAG Benchmark.}  
Retrieval NDCG@5 results of Rerankers after first step retrieval with ColPali and ColQwen.
}
\label{Table:real-mm-rag}
\vspace{-0.4cm}
\end{table}

%%%%%%%%%%%%%%%%%%%%%%%%%%%%%%%%%%%%%%%%%%%
% Results
%%%%%%%%%%%%%%%%%%%%%%%%%%%%%%%%%%%%%%%%%%
\section{Results}

In this section, we demonstrate the effectiveness of our data generation framework for reranker training. We first describe the experimental setup in \cref{sec:experimental_setup}, then show in \cref{sec:main_results} that training with our generated data significantly outperforms strong baselines under comparable settings. We further show in \cref{sec:main_results} that combining domain-specific data targeting reranker weaknesses and rephrased queries leads to additional improvements. Finally, we present in \cref{sec:ablations} ablations to assess the contribution of each dataset and compare against traditional document-level hard negative mining.

\subsection{Experimental Setup} \label{sec:experimental_setup}

\paragraph{Benchmarks.} 
We evaluate on two multimodal retrieval benchmarks that closely reflect real-world RAG use cases, featuring challenging and information-seeking queries.
\emph{\textbf{ViDoReV2}}~\citep{vidore_benchmark_v2}: This benchmark includes 8 evaluation datasets, three of which are multilingual. Some queries have answers that span multiple pages, contributing to overall lower performance.
\emph{\textbf{Real-MM-RAG}}
~\citep{wasserman2025real}: This benchmark includes four high-difficulty evaluation set: FinReport, FinSlides, TechReport, and TechSlides. We also evaluate on the rephrased version of this benchmark, provided by the authors, to assess model robustness and true semantic understanding.

\paragraph{Evaluation Metric} 
We report the standard NDCG@5 as the primary evaluation metric, measuring the quality of the top-ranked retrieved pages. Additional metrics, such as Recall@5 and NDCG@10, are reported in the Appendix (see {\cref{sup_sec:additional_results}}).

\paragraph{Retrieval Models.}
To evaluate the reranker’s impact, we use two strong retrieval models following the ColPali paper~\citep{faysse2024colpali} approach: \emph{\textbf{ColPali-v1.2}} and \emph{\textbf{ColQwen2-v1.0}}.
We first retrieved top-20 pages per each query in the evaluation dataset using those models and then used the rerankers for reordering those top-20 pages.

\paragraph{Baseline Rerankers.}
As multimodal reranking is still a developing field, we compare against the following strong and relevant baselines: \emph{\textbf{Qwen-VLM}} uses the Qwen2-VL-2B-Instruct model with our standard reranking prompt but without any fine-tuning. \emph{\textbf{MonoQwen}} is a fine-tuned reranker trained using the MonoQwen approach on the ColPali dataset. It uses the same base model (Qwen2-VL-2B-Instruct) and training objective as our reranker but is trained solely on hard-negative-mined document-level data, without query generation or adaptation to model weaknesses.

%%%%%%%%%%%%%%%%%%%%%%%%%%%%%%%%%%%%%%%%%%%%%%%%%%%

\begin{table}[b]
\footnotesize
\renewcommand{\arraystretch}{1.5} 
\setlength{\dashlinedash}{0.5pt}
\setlength{\dashlinegap}{0.5pt}
\setlength\tabcolsep{1.4pt} 
\hspace{-0.25cm}
\begin{tabular}{@{\extracolsep{\fill}}lccccc} 
\toprule
\multirow{2}{*}{\textbf{Benchmark}} 
  & \textbf{FinReport} & \textbf{FinSlides} & \textbf{TechReport} & \textbf{TechSlides} 
  & \multirow{2}{*}{\textbf{Avg}} \\
  & \textbf{Rephrased} & \textbf{Rephrased} & \textbf{Rephrased}  & \textbf{Rephrased}  & \\
\midrule

\textit{ColQwen}   
  & 41.8 & 31.1 & 67.2 & 78.0 & 54.5 \\

\hspace{0.1cm}\textit{Qwen-VLM}  
  & 49.3 {\tiny \textcolor{DarkGreen}{+7.5}}  
  & 49.0 {\tiny \textcolor{DarkGreen}{+17.9}}  
  & 60.6 {\tiny \textcolor{red}{–6.6}}  
  & 73.7 {\tiny \textcolor{red}{–4.3}}  
  & 58.2 {\tiny \textcolor{DarkGreen}{+3.7}}\\

\hspace{0.1cm}\textit{MonoQwen}  
  & 49.0 {\tiny \textcolor{DarkGreen}{+7.2}}  
  & 50.7 {\tiny \textcolor{DarkGreen}{+19.6}}  
  & 73.0 {\tiny \textcolor{DarkGreen}{+5.8}}  
  & 82.6 {\tiny \textcolor{DarkGreen}{+4.6}}  
  & 63.8 {\tiny \textcolor{DarkGreen}{+9.3}}\\

\hspace{0.1cm}\textit{\textbf{DocReRank}-F \raisebox{-1.8ex}{\hspace{-3.1em}\tiny (w/o Reph)}}  
  & \textbf{55.0} {\tiny \textcolor{DarkGreen}{+13.2}}  
  & \underline{\textbf{53.1}} {\tiny \textcolor{DarkGreen}{+22.0}}  
  & \textbf{72.5} {\tiny \textcolor{DarkGreen}{+5.3}}  
  & \textbf{83.1} {\tiny \textcolor{DarkGreen}{+5.1}}  
  & \textbf{65.9} {\tiny \textcolor{DarkGreen}{+11.4}}\\

\hspace{0.1cm}\textit{\textbf{DocReRank}-F}  
  & \underline{\textbf{57.1}} {\tiny \textcolor{DarkGreen}{+15.3}}  
  & \textbf{52.1} {\tiny \textcolor{DarkGreen}{+21.0}}  
  & \underline{\textbf{79.0}} {\tiny \textcolor{DarkGreen}{+11.8}}  
  & \underline{\textbf{88.0}} {\tiny \textcolor{DarkGreen}{+10.0}}  
  & \underline{\textbf{69.0}} {\tiny \textcolor{DarkGreen}{+14.5}}\\

\bottomrule
\end{tabular}
\caption{
\textbf{Performance on the Rephrased Real-MM-RAG Benchmark.}  
Retrieval NDCG@5 results of Rerankers after first step retrieval with ColQwen.}
\label{Table:real-mm-rag_reph}
\vspace{-0.4cm}
\end{table}

\begin{table*}[ht!]
\footnotesize
\renewcommand{\arraystretch}{1.5} % Adjust row height
\setlength{\dashlinedash}{0.5pt}
\setlength{\dashlinegap}{0.5pt}
\setlength\tabcolsep{1.5pt} % Adjust column spacing
% \hspace{-0.15cm}
\begin{tabular}{@{\extracolsep{\fill}}lccccccccc} 
\toprule
\textbf{Benchmark} & \textbf{Axa} & \textbf{Economics} & \textbf{Restaurant-rse} & \textbf{Restaurant-esg}  & \textbf{Biomedical} & \textbf{Economics-ML} & \textbf{Restaurant-ML} & \textbf{Biomedical-ML} & \textbf{Avg}\\
\midrule
\textit{ColQwen}   & 64.4 & 61.3 & 54.4 & 61.2 & 62.2 & 52.6 & 56.0 & 56.8 & 58.6 \\

\hspace{0.05cm}\textit{Qwen-VLM}   
  & 66.1 {\tiny \textcolor{DarkGreen}{+1.7}}
  & 51.9 {\tiny \textcolor{red}{–9.4}}
  & 63.6 {\tiny \textcolor{DarkGreen}{+9.2}}
  & 64.2 {\tiny \textcolor{DarkGreen}{+3.0}}
  & 57.7 {\tiny \textcolor{red}{–4.5}}
  & 50.4 {\tiny \textcolor{red}{–2.2}}
  & 62.9 {\tiny \textcolor{DarkGreen}{+6.9}}
  & 56.3 {\tiny \textcolor{red}{–0.5}}
  & 59.1 {\tiny \textcolor{DarkGreen}{+0.5}}\\

\hspace{0.1cm}\textit{FT on Col-HNDoc}   
  & 73.2 {\tiny \textcolor{DarkGreen}{+8.8}}
  & 64.4 {\tiny \textcolor{DarkGreen}{+3.1}}
  & 62.3 {\tiny \textcolor{DarkGreen}{+7.9}}
  & 66.1 {\tiny \textcolor{DarkGreen}{+4.9}}
  & 67.8 {\tiny \textcolor{DarkGreen}{+5.6}}
  & 56.5 {\tiny \textcolor{DarkGreen}{+3.9}}
  & 62.3 {\tiny \textcolor{DarkGreen}{+6.3}}
  & 62.8 {\tiny \textcolor{DarkGreen}{+6.0}}
  & 64.4 {\tiny \textcolor{DarkGreen}{+5.8}}\\

\hspace{0.05cm}\textit{\textbf{DocReRank}-Base}   
  & 77.3 {\tiny \textcolor{DarkGreen}{+12.9}}
  & 67.5 {\tiny \textcolor{DarkGreen}{+6.2}}
  & 63.3 {\tiny \textcolor{DarkGreen}{+8.9}}
  & 71.7 {\tiny \textcolor{DarkGreen}{+10.5}}
  & 68.5 {\tiny \textcolor{DarkGreen}{+6.3}}
  & 60.7 {\tiny \textcolor{DarkGreen}{+8.1}}
  & 65.2 {\tiny \textcolor{DarkGreen}{+9.2}}
  & 63.8 {\tiny \textcolor{DarkGreen}{+7.0}}
  & 67.2 {\tiny \textcolor{DarkGreen}{+8.6}}\\

\hspace{0.05cm}\textit{\textbf{DocReRank}-B \tiny{w Fin}}   
  & \underline{\textbf{76.8}} {\tiny \textcolor{DarkGreen}{+12.4}}
  & 66.8 {\tiny \textcolor{DarkGreen}{+5.5}}
  & \textbf{67.5} {\tiny \textcolor{DarkGreen}{+13.1}}
  & 73.8 {\tiny \textcolor{DarkGreen}{+12.6}}
  & 60.2 {\tiny \textcolor{red}{–2.0}}
  & \underline{\textbf{68.9}} {\tiny \textcolor{DarkGreen}{+16.3}}
  & \textbf{66.5} {\tiny \textcolor{DarkGreen}{+10.5}}
  & 64.3 {\tiny \textcolor{DarkGreen}{+7.5}}
  & 68.1 {\tiny \textcolor{DarkGreen}{+9.5}}\\

\hspace{0.05cm}\textit{\textbf{DocReRank}-B \tiny{w Fin\&Reph}}   
  & 73.6 {\tiny \textcolor{DarkGreen}{+9.2}}
  & \textbf{67.9} {\tiny \textcolor{DarkGreen}{+6.6}}
  & \underline{\textbf{68.1}} {\tiny \textcolor{DarkGreen}{+13.7}}
  & \textbf{74.8} {\tiny \textcolor{DarkGreen}{+13.6}}
  & \textbf{70.3} {\tiny \textcolor{DarkGreen}{+8.1}}
  & 62.9 {\tiny \textcolor{DarkGreen}{+10.3}}
  & \underline{\textbf{67.0}} {\tiny \textcolor{DarkGreen}{+11.0}}
  & \underline{\textbf{65.5}} {\tiny \textcolor{DarkGreen}{+8.7}}
  & \underline{\textbf{68.8}} {\tiny \textcolor{DarkGreen}{+10.2}}\\

\hspace{0.05cm}\textit{\textbf{DocReRank}-Full}   
  & \textbf{75.3} {\tiny \textcolor{DarkGreen}{+10.9}}
  & \underline{\textbf{68.7}} {\tiny \textcolor{DarkGreen}{+7.4}}
  & 66.6 {\tiny \textcolor{DarkGreen}{+12.2}}
  & \underline{\textbf{75.8}} {\tiny \textcolor{DarkGreen}{+14.6}}
  & \textbf{68.2} {\tiny \textcolor{DarkGreen}{+6.0}}
  & \textbf{64.7} {\tiny \textcolor{DarkGreen}{+12.1}}
  & 66.0 {\tiny \textcolor{DarkGreen}{+10.0}}
  & \textbf{64.7} {\tiny \textcolor{DarkGreen}{+7.9}}
  & \underline{\textbf{68.8}} {\tiny \textcolor{DarkGreen}{+10.2}}\\

\bottomrule
\end{tabular}
\vspace{-0.1cm}
\caption{
\textbf{Ablation Results on ViDoReV2 Benchmark.}  
Retrieval NDCG@5 Results.  
We compare a model fine-tuned only on document-based hard negatives (\emph{FT on Col-HNDoc})
% \emph{FT on Col-HNDoc}, which serves as the baseline model fine-tuned only on document-based hard negatives, 
to our \emph{DocReRank-Base}, which outperforms this baseline. Adding finance-, fine-detail-specific generated queries (\emph{Fin-HNQue}), and rephrased data leads to further performance gains.
}
\label{Table:ablation}
\vspace{-0.4cm}
\end{table*}

\subsection{Main Results} \label{sec:main_results}

In \cref{Table:vidoreV2,Table:real-mm-rag}, we report NDCG@5 reranking results after retrieving with both ColPali and ColQwen. Retrieval-only performance is also shown for reference.
We first evaluate our base model, \emph{\textbf{DocReRank-Base}}, which fine-tunes Qwen2-VL using a combined training set: half with negative pages from traditional document-level hard negative mining (\emph{Col-HNDoc}), and half with hard negative queries from our generation approach (\emph{Col-HNQue}). This dataset includes 120K positive examples and 360K negatives. This setup allows a direct comparison to MonoQwen, which uses a similar architecture and training data but relies only on document-based negatives. 
 
 As shown in the results, \emph{DocReRank-Base} achieves significant improvements over retrieval-only baselines (e.g., +8.6 points with ColQwen on ViDoReV2), and clearly outperforms MonoQwen, which as far as we know, differs only on the training data (document hard negatives only). On ColQwen, we observe gains of +2.8 on ViDoReV2 and +4.7 on Real-MM-RAG.
 
We also evaluate our full model, \emph{DocReRank-Full}, which incorporates the Finance-Focused Generation dataset (\emph{Fin-HNQue}) and rephrased positive queries. This leads to additional gains across both retrieval models and benchmarks, demonstrating the impact of adapting to model-specific weaknesses and requiring model sematic understanding.

To further highlight the role of rephrasing, we evaluate on the rephrased version of Real-MM-RAG. In \cref{Table:real-mm-rag_reph}, we compare our full model trained with and without rephrased positives. Results show that training with rephrased queries improves robustness, although some performance drop remains when evaluated on rephrased benchmarks—emphasizing the challenge of moving beyond shallow keyword matching toward true semantic understanding.

%%%%%%%%%%%%%%%%%%%%%%%%%%%%%%%%%%%%%%%%%%%%%%%%%%%

\subsection{Ablation Studies}   \label{sec:ablations}

In this subsection, we aim to demonstrate two things:
(i) our data generation approach offers clear benefits over using only document-based hard negatives, and
(ii) the individual contribution of each of our generated datasets. 

To isolate the impact of our data, we fine-tuned the same model used in \emph{DocReRank}, under identical training settings, but only with the \emph{Col-HNDoc} dataset (based on document hard negative retreival). As expected, this model achieved results similar to MonoQwen, which was trained in a comparable manner. As shown in \cref{Table:ablation}, our \emph{DocReRank-Base} model outperforms the model trained solely on \emph{Col-HNDoc}, demonstrating the added value of our query generation approach. Adding the \emph{Fin-HNQue} (Finance Hard Negative Queries) dataset leads to further improvements, and incorporating the rephrased dataset boosts performance even more.

Importantly, all \emph{DocReRank-Base} models were trained using the same number of total training examples, sampled from different datasets (see \cref{sup_sec:training_detials} for details). The full model was trained with twice the number of examples. While it shows comparable results in this table, it achieved additional improvements with ColPali retrieval (see \cref{Table:ablation_colpali}).

%%%%%%%%%%%%%%%%%%%%%%%%%%%%%%%%%%%%%%%%%%%%%%%%%%%
\section{Conclusions}

Our work challenges the conventional reliance on document-level hard negative mining for reranker training by introducing a query-generation alternative. A core insight is that generation offers greater controllability and diversity than retrieval from a fixed document corpus. Query generation is also more practical, as queries are short and easy to control. Grounding query generation in a single document page, gives even more controllability—enabling generation of multiple, diverse negatives tailored to specific content and allowing efficient verification. This enables us to generate harder negatives, verify unanswerability, and target known model weaknesses.  We further show how this controllability can be used to generate specific negatives that match model-specific weaknesses. It can also be adapted to application-specific needs. For example, ensuring a model distinguishes machine type when answering questions about manufacturing manuals, to avoid returning answers for the wrong machine. 

Our results show that query-level generation is a strong alternative to document mining, yielding superior reranking performance when used alongside traditional negatives. We believe this framework can be extended and refined to provide valuable training data for future research and deployment.

%%%%%%%%%%%%%%%%%%%%%%%%%%%%%%%%%%%%%%%%%%%%%%%%%%%
\section{Limitations}

While our approach represents a significant step toward better hard negative examples and has been shown to improve reranker training, several limitations remain.
% While DocReRank represents a significant step towards more robust document reranking for RAG, several limitations remain.}
\emph{Query variability:} Positive and corresponding hard negative queries are generated using a pipeline of VLMs and LLMs. Despite careful prompt instructions, not the full query space used by a human might be exploited. 
\emph{Query verification:} To verify the answerability of a given query, VLMs are used. Nevertheless, despite strategies such as double verification using two separate prompts, false negatives and positives  can occur, potentially limiting quality of hard negatives. 
\emph{Reranker Dependency:} The reranker step fully depends on the initially provided ranked subset of the full document corpus by the retrieval algorithm. If the true positive document of a query is not listed in the provided subset, the reranker will never be able to provide the true answer neither.

% Custom bibliography entries only
\bibliography{custom}

\appendix

\clearpage
\section{Appendix}
\label{sec:appendix}

\renewcommand{\thefigure}{S\arabic{figure}} % Figures as S1, S2, ...
\renewcommand{\thetable}{S\arabic{table}}   % Tables as S1, S2, ...
\setcounter{figure}{0}  % Reset numbering for appendix
\setcounter{table}{0}

\subsection{Evaluation Details}
\label{sup_sec:Evaluation_details}

All evaluations—both for retrieval models and rerankers—were conducted using the ColPali evaluation framework and the \texttt{mteb} package. For reranker evaluation, we first retrieved the top-20 pages per query using a given retrieval model. Then, for each query, the reranker computed a relevance score between the query and each of the top-20 retrieved pages. These pages were subsequently re-ordered according to the reranker's relevance scores, and evaluation metrics were computed on the newly ranked list.

\subsection{Training Details}
\label{sup_sec:training_detials}

Our models were trained using data from the following datasets:
(i) \emph{\textbf{Col-HNDoc}} – document-level hard negative mining based on the ColPali training set;
(ii) \emph{\textbf{Col-HNQue}} – our hard negative query generation applied to the ColPali training set;
(iii) \emph{\textbf{Fin-HNQue}} – our finance-specific hard negative query dataset;
(iv) \emph{\textbf{Reph-Col-HNDoc}} and
(v) \emph{\textbf{Reph-Col-HNQue}} – rephrased variants of the above datasets.
\\
\noindent
Below we detail the data used for training each model. For simplicity, we report the number of positive examples used (each paired with 3 negatives, totaling 4× the size in training samples).

\begin{itemize}[leftmargin=*, itemsep=2pt, topsep=2pt, parsep=0pt]
\item \emph{\textbf{FT on Col-HNDoc}} – trained on the full 120k positives and 360k negatives from \emph{Col-HNDoc} using document-based hard negatives.

\item \emph{\textbf{DocReRank-Base}} – trained on 60k positives from \emph{Col-HNDoc} and 60k from \emph{Col-HNQue}.

\item \emph{\textbf{DocReRank-B w/ Fin}} – trained on 60k from \emph{Col-HNDoc}, 40k from \emph{Col-HNQue}, and the full 20k from \emph{Fin-HNQue}.

\item \emph{\textbf{DocReRank-B w/ Fin \& Reph}} – trained on 60k from \emph{Reph-Col-HNDoc}, 40k from \emph{Reph-Col-HNQue}, and 20k from \emph{Fin-HNQue}.

\item \emph{\textbf{DocReRank-Full}} – trained on 120k from \emph{Reph-Col-HNDoc}, 120k from \emph{Reph-Col-HNQue}, and 20k from \emph{Fin-HNQue}.
\end{itemize}

\subsection{Additional Results} 
\label{sup_sec:additional_results}

In the main paper, we focused on reporting the NDCG@5 metric. In \cref{Table:real-mm-rag_recall1,Table:real-mm-rag_recall5,Table:real-mm-rag_ndcg10,Table:vidoreV2_recall1,Table:vidoreV2_recall5,Table:vidoreV2_ncdg10}, we provide results for additional metrics, including Recall@1, Recall@5, and NDCG@10. We also report additional ablation results in \cref{Table:ablation_colpali}, using ColPali retrieval, complementing the ColQwen-based ablations shown in \cref{Table:ablation}.

\subsection{Generation and Verification Prompts}
\begin{figure*}[h!]
    \centering
    \includegraphics[width=1.0 \textwidth]
    {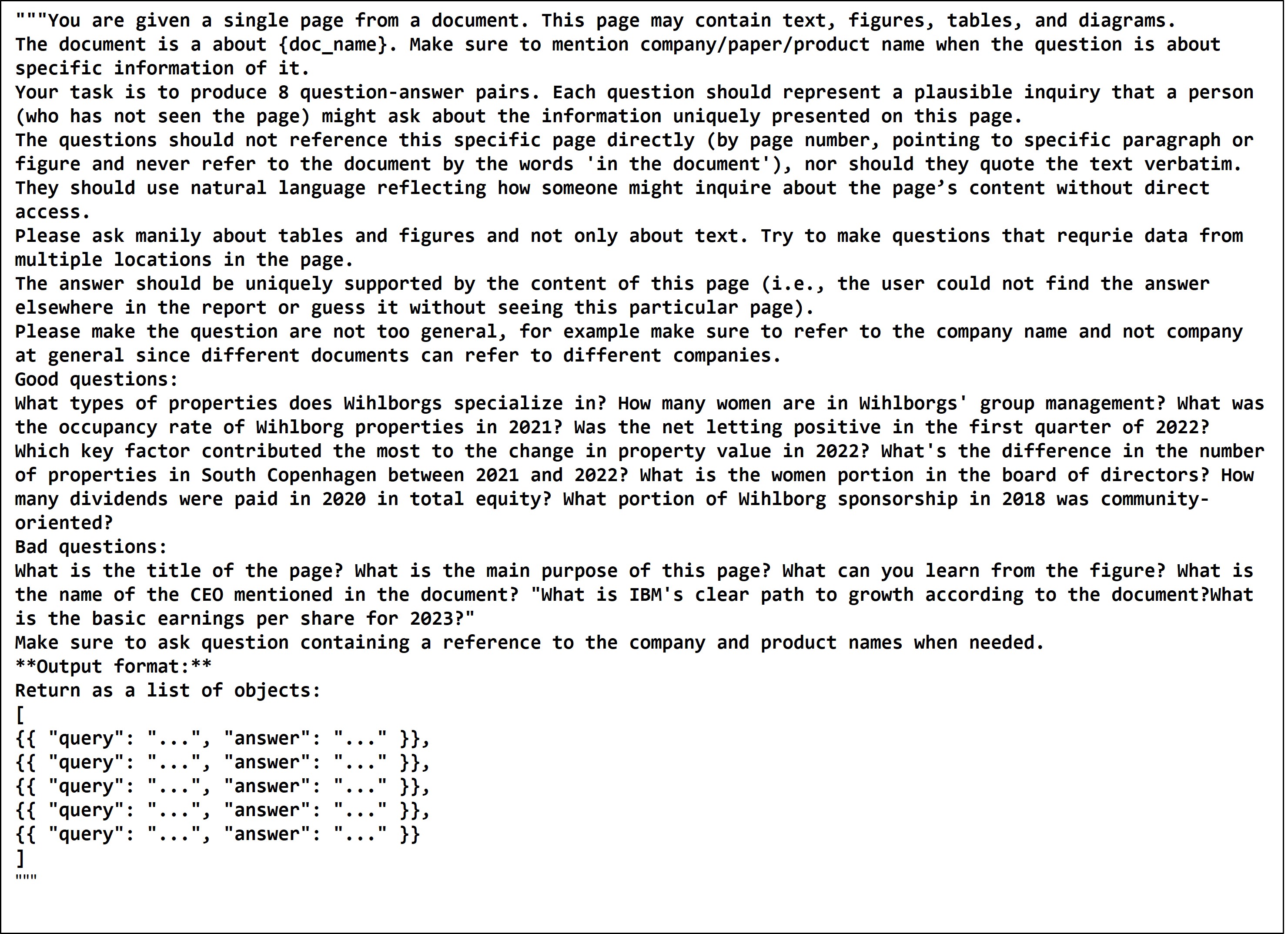} 
    \vspace{-0.7cm}
    \caption{\textbf{Positive Query Generation Prompt:} Creating RAG style queries, answerable by the corresponding document, using a Pixtral-12B VLM \citet{agrawal2024pixtral}. $N$ positive candidates are generated with the given prompt. The prompt emphasizes multimodal understanding by focusing on page elements such as figures, tables, and diagrams.}
    \vspace{-0.2cm}
    \label{fig_sup:pos_gen_prompt}
\end{figure*}

\begin{figure*}[h!]
    \centering
    \includegraphics[width=1.0\textwidth]{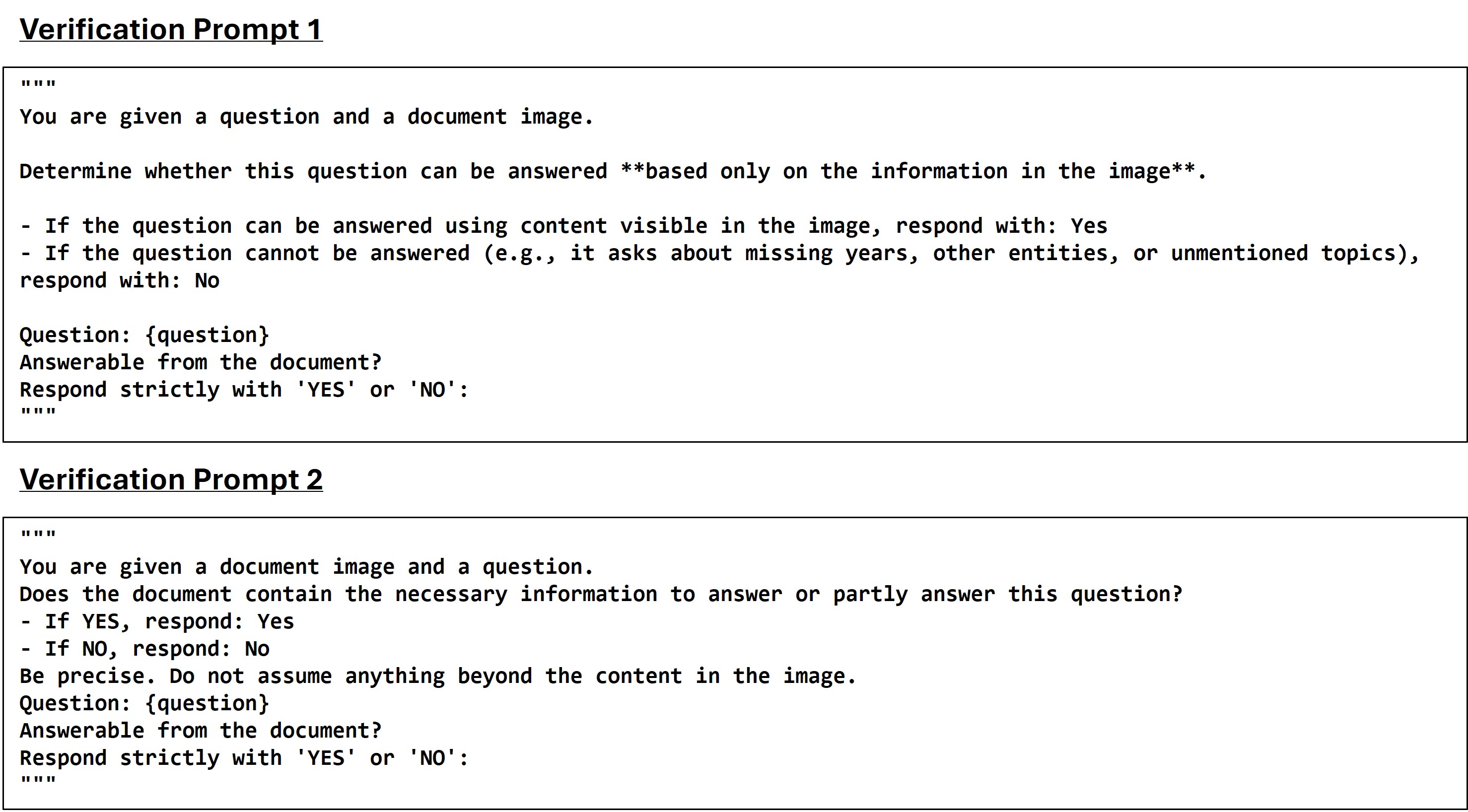} 
    \vspace{-0.5cm}
    \caption{\textbf{Query Verification Prompt.} Verifying whether a query is answerable from the page content using the Qwen2.5-VL-7B-Instruct VLM. Two slightly different prompts are used to improve verification robustness. For positive queries, only those marked as answerable by both prompts are kept; for negatives, any query marked as answerable by either prompt is filtered out.}
    \vspace{-0.2cm}
    \label{fig_sup:verif_prompts}
\end{figure*}

\begin{figure*}[h!]
    \centering
    \includegraphics[width=1.\textwidth]{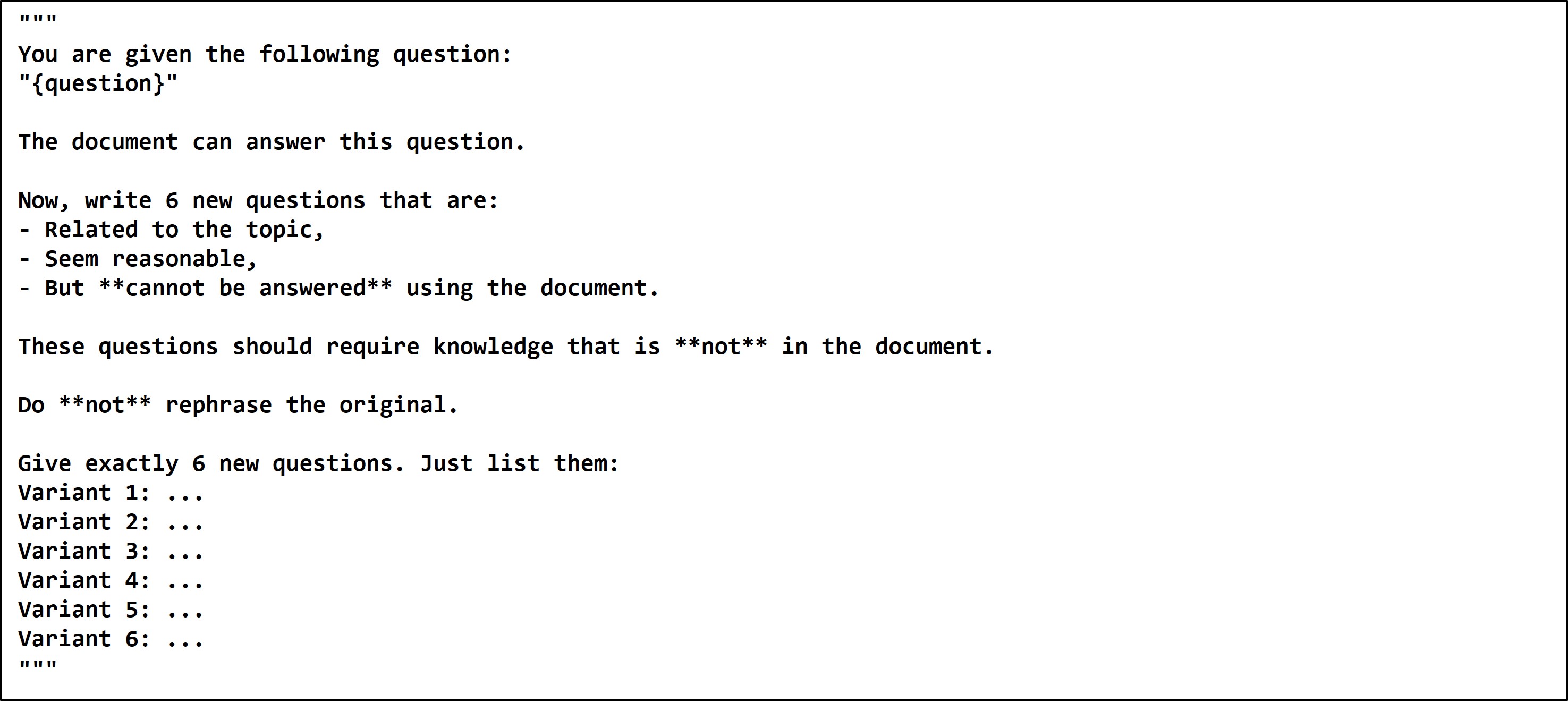} 
    \caption{\textbf{Negative Generation Prompt:} Given a positive query and the corresponding document, hard negative queries - queries that are unanswerable by the current document - are created. As it is relatively easy for an LLM to generate semantically close variants, Qwen2.5-7B-Instruc LLM is used to generate 12 negative variants of the positive query. The prompt is focused to create negatives similar in topic and form, but different in information.}
    \vspace{-0.2cm}
    \label{fig_sup:neg_gen_prompt}
\end{figure*}

\begin{figure*}[h!]
    \centering
    \includegraphics[width=1.0\textwidth]{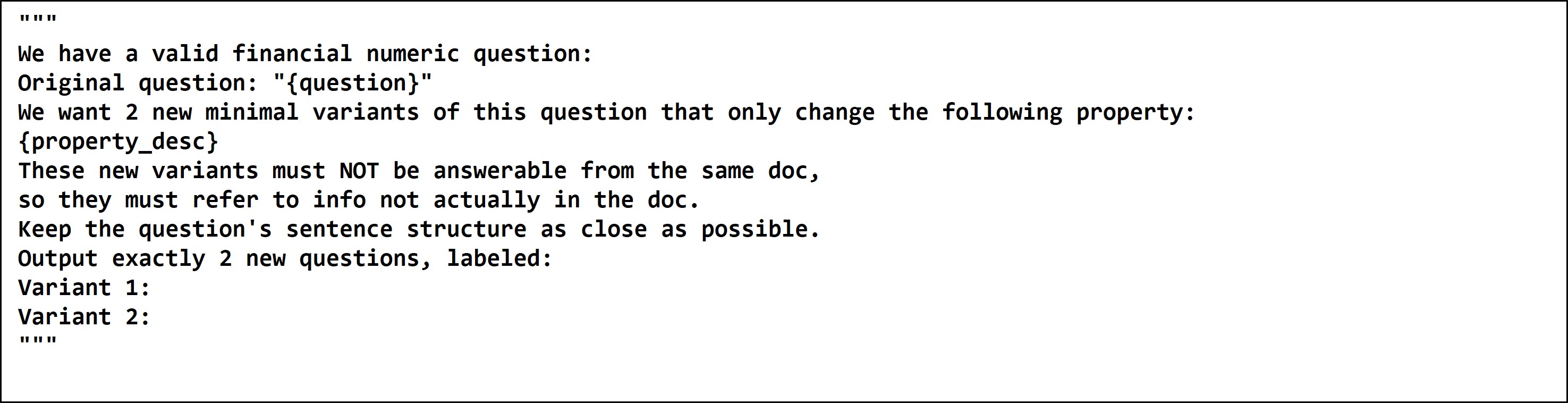} 
    \caption{\textbf{Finance Negative Generation Prompt:} To handle fine-grained information distinctions that occur e.g.\ in financial reports, a dedicated prompt has been created. A dedicated set of prompts has been developed to generate a variant by modifying exactly one property \texttt{\{property\_desc\}}, such as the year (e.g., 2022 → 2024), company name (e.g., Apple → IBM), numerical value (e.g., price, percentage), financial metric (e.g., revenue, sales, acquisitions), subject metric (e.g., dividends, stocks, options), or business segment (e.g., cloud, software, manufacturing).}
    \vspace{-0.2cm}
    \label{fig_sup:fin_neg_gen_prompt}
\end{figure*}

Our multi-step query generation pipeline combines a Large Language Model (LLM) and a Vision-Language Model (VLM), each guided by specific prompts tailored to different stages of the process. We provide here the full prompts used in each step. The positive query generation prompt, shown in \cref{fig_sup:pos_gen_prompt}, is used with the VLM (Pixtral-12B) to generate natural, information-seeking queries that are answerable from the given document page. The two verification prompts, shown in \cref{fig_sup:verif_prompts}, are used with the VLM (Qwen2.5-VL-7B-Instruct) to determine whether a query is answerable from the page content. Using two slightly different prompt formulations improves verification robustness. The generic hard negative query generation prompt, shown in \cref{fig_sup:neg_gen_prompt}, instructs the LLM to rephrase a given positive query into unanswerable variants that are similar in form and topic. Finally, the finance-specific prompt, shown in \cref{fig_sup:fin_neg_gen_prompt}, focuses on fine-grained factual attributes (e.g., years, amounts, company names) to create particularly challenging negative queries for detail-sensitive content. 

\vspace{-0.2cm}
\subsection{Licensing and General Information}

All models and datasets used in this work comply with their respective licenses. Qwen2-VL (ColQwen2) and Qwen are licensed under Apache 2.0, with adapters released under the MIT license. PaliGemma (ColPali) follows the Gemma license, also with adapters under MIT. Pixtral-12B-2409 (mistralai) and Mixtral-8x22B are both released under the Apache 2.0 license, which permits unrestricted use, modification, and distribution. LLaMA 3.3 70B is released under the LLaMA 3.3 Community License Agreement.

All datasets used are in English, except for ViDoRe V2, which includes queries and documents in French. The ColPali training set includes subsampled academic datasets redistributed under their original licenses. It also incorporates synthetic datasets generated from publicly available internet content and VLM-generated queries, which are released without usage restrictions. The REAL-MM-RAG benchmark is distributed under the Community Data License Agreement – Permissive, Version 2.0 (CDLA-Permissive-2.0). The FinTabNet dataset is composed of data collected from publicly available sources. An AI assistant (ChatGPT) was used for minor grammar and sentence structure edits.

\centering
\begin{table*}[ht!]
\footnotesize
\renewcommand{\arraystretch}{1.5} % Adjust row height
\setlength{\dashlinedash}{0.5pt}
\setlength{\dashlinegap}{0.5pt}
\setlength\tabcolsep{2.5pt} % Adjust column spacing
\hspace{-0.15cm}
\begin{tabular}{@{\extracolsep{\fill}}lccccccccc} 
\toprule
\textbf{Benchmark} & \textbf{Axa} & \textbf{Economics} & \textbf{Restaurant-rse} & \textbf{Restaurant-esg}  & \textbf{Biomedical} & \textbf{Economics-ML} & \textbf{Restaurant-ML} & \textbf{Biomedical-ML} & \textbf{Avg} \\
\midrule

\textit{ColPali}  & 18.4               & 8.7  & 21.5  & 38.3               & 32.6  & 9.0   & 22.1  & 32.0               & 22.8  \\

\textit{Qwen-VLM} & \textbf{36.3}      & 6.0  & 27.1  & 35.6               & 30.5  & 7.1   & 26.3  & 29.8               & 24.8  \\

\textit{MonoQwen} & 31.8               & 9.4  & 27.4  & \textbf{49.7}      & 37.0  & 8.5   & 26.5  & 36.5               & 28.4  \\

\textit{\textbf{DocReRank}-Base} & \underline{\textbf{37.2}} & \underline{\textbf{13.9}} & \textbf{30.4} & \underline{\textbf{50.5}} & \underline{\textbf{40.9}} & \textbf{11.5} & \textbf{32.6} & \underline{\textbf{37.6}} & \textbf{31.8} \\

\textit{\textbf{DocReRank}-Full} & \underline{\textbf{37.2}} & \textbf{12.8}            & \underline{\textbf{33.6}} & 48.6               & \textbf{39.2}            & \underline{\textbf{13.0}} & \underline{\textbf{33.3}} & \textbf{37.2}            & \underline{\textbf{31.9}} \\

\addlinespace

\textit{ColQwen}  & 29.1 & 5.9  & 22.3 & 41.9 & 36.7 & 6.4  & 24.6 & 33.1 & 25.0 \\

\textit{Qwen-VLM} & \underline{\textbf{37.2}} & 7.6  & 27.7 & 34.3 & 30.1 & 7.5  & 27.8 & 29.9 & 25.3 \\

\textit{MonoQwen} & \textbf{31.8} & 9.5  & 28.1 & 48.4 & 36.8 & 9.1  & 26.7 & 36.9 & 28.4 \\

\textit{\textbf{DocReRank}-Base}  & \underline{\textbf{37.2}} & \underline{\textbf{13.1}} & \underline{\textbf{33.5}} & \textbf{49.1} & \underline{\textbf{40.9}} & \textbf{11.4} & \underline{\textbf{34.6}} & \underline{\textbf{38.2}} & \underline{\textbf{32.2}} \\

\textit{\textbf{DocReRank}-Full}  & \underline{\textbf{37.2}} & \textbf{12.3} & \textbf{33.3} & \underline{\textbf{50.2}} & \textbf{38.7} & \underline{\textbf{13.2}} & \textbf{33.0} & \textbf{37.5} & \textbf{31.9} \\

\bottomrule
\end{tabular}
\caption{
\textbf{Performance on the ViDoReV2
Benchmark recall@1}  
}
\label{Table:vidoreV2_recall1}
\vspace{-0.4cm}
\end{table*}

\vspace{1em}

\begin{table*}[ht!]
\footnotesize
\renewcommand{\arraystretch}{1.5} % Adjust row height
\setlength{\dashlinedash}{0.5pt}
\setlength{\dashlinegap}{0.5pt}
\setlength\tabcolsep{2.5pt} % Adjust column spacing
\hspace{-0.15cm}
\begin{tabular}{@{\extracolsep{\fill}}lccccccccc} 
\toprule
\textbf{Benchmark} & \textbf{Axa} & \textbf{Economics} & \textbf{Restaurant-rse} & \textbf{Restaurant-esg}  & \textbf{Biomedical} & \textbf{Economics-ML} & \textbf{Restaurant-ML} & \textbf{Biomedical-ML} & \textbf{Avg} \\
\midrule

\textit{ColPali} & 58.5                     & 27.1                     & 54.6                     & 60.0                     & 61.3                     & 24.3                     & 56.5                     & 58.7                     & 50.1 \\

\textit{Qwen-VLM} & 58.0                     & 27.2                     & \textbf{59.6}            & 66.9                     & 59.9                     & 26.9                     & \textbf{61.6}            & 59.7                     & 52.1 \\

\textit{MonoQwen} & \textbf{64.6}            & 30.4                     & 59.0                     & 69.1                     & \underline{\textbf{68.6}} & 31.5                     & \underline{\textbf{61.7}}                     & \textbf{64.3}            & 56.0 \\

\textit{\textbf{DocReRank}-Base} & 64.3                     & \textbf{34.0}            & 58.2                     & \textbf{70.3}            & \textbf{67.9}            & \textbf{30.5}            & 60.7                     & 63.4                     & \textbf{56.2} \\

\textit{\textbf{DocReRank}-Full} & \underline{\textbf{65.1}} & \underline{\textbf{36.1}} & \underline{\textbf{60.4}} & \underline{\textbf{72.3}} & 67.4                     & \underline{\textbf{32.6}} & \textbf{61.6}            & \underline{\textbf{65.0}} & \underline{\textbf{57.6}} \\

\addlinespace

\textit{ColQwen} & 59.0                      & \textbf{35.2}             & 56.6                     & 66.1                     & 64.0                     & 29.3                     & 58.4                     & 59.4                     & 53.5                     \\

\textit{Qwen-VLM} & 58.0                      & 27.2                      & \underline{\textbf{67.2}} & 71.1                     & 62.2                     & 26.9                     & \underline{\textbf{66.2}} & 60.4                     & 54.9                     \\

\textit{MonoQwen} & 66.5                      & 31.6                      & 64.6                     & \textbf{73.6}            & \underline{\textbf{70.2}} & 31.5                     & 65.7                     & \textbf{64.9}            & \textbf{58.6}            \\

\textit{\textbf{DocReRank}-Base} & \underline{\textbf{68.7}} & \underline{\textbf{35.8}} & 61.8                     & 72.5                     & 68.6                     & \textbf{31.8}            & 63.9                     & 64.6                     & 58.5                     \\

\textit{\textbf{DocReRank}-Full} & \textbf{67.8}            & \underline{\textbf{35.8}} & \textbf{66.1}            & \underline{\textbf{77.6}} & \textbf{69.4}            & \underline{\textbf{34.1}} & \textbf{66.1}            & \underline{\textbf{66.3}} & \underline{\textbf{60.4}} \\

\bottomrule
\end{tabular}
\caption{
\textbf{Performance on the ViDoReV2
Benchmark recall@5}  
}
\label{Table:vidoreV2_recall5}
\vspace{-0.4cm}
\end{table*}

\vspace{1em}

\begin{table*}[ht!]
\footnotesize
\renewcommand{\arraystretch}{1.5} % Adjust row height
\setlength{\dashlinedash}{0.5pt}
\setlength{\dashlinegap}{0.5pt}
\setlength\tabcolsep{2.5pt} % Adjust column spacing
\hspace{-0.15cm}
\begin{tabular}{@{\extracolsep{\fill}}lccccccccc} 
\toprule
\textbf{Benchmark} & \textbf{Axa} & \textbf{Economics} & \textbf{Restaurant-rse} & \textbf{Restaurant-esg}  & \textbf{Biomedical} & \textbf{Economics-ML} & \textbf{Restaurant-ML} & \textbf{Biomedical-ML} & \textbf{Avg} \\
\midrule

\textit{ColPali}  & 54.7 & 52.2 & 54.7 & 58.9 & 61.5 & 47.5 & 55.9 & 59.0 & 55.6 \\

\textit{Qwen-VLM}  & 64.2 & 49.4 & 60.8 & 62.5 & 60.2 & 50.2 & 61.5 & 59.2 & 58.5 \\

\textit{MonoQwen}  & 64.5 & 55.8 & \textbf{61.8} & \textbf{70.6} & 68.5 & 54.0 & 62.5 & \textbf{65.7} & 62.9 \\

\textit{\textbf{DocReRank}-Base}  & \underline{\textbf{69.6}} & \textbf{61.5} & 60.2 & 70.1 & \underline{\textbf{69.5}} & \textbf{56.1} & \textbf{64.2} & 65.2 & \textbf{64.6} \\

\textit{\textbf{DocReRank}-Full}  & \textbf{67.6} & \underline{\textbf{64.2}} & \underline{\textbf{62.0}} & \underline{\textbf{72.3}} & \textbf{69.0} & \underline{\textbf{59.5}} & \underline{\textbf{65.9}} & \underline{\textbf{67.0}} & \underline{\textbf{66.0}} \\

\addlinespace

\textit{ColQwen}  & 66.9 & 58.2 & 59.6 & 63.8 & 66.1 & 51.3 & 60.7 & 60.7 & 60.9 \\

\textit{Qwen-VLM}  & 70.1 & 54.3 & 67.1 & 66.8 & 61.9 & 51.3 & 66.8 & 59.8 & 62.3 \\

\textit{MonoQwen}  & 71.0 & 60.3 & 65.5 & \textbf{74.8} & 70.0 & 57.1 & 65.9 & \textbf{66.6} & 66.4 \\

\textit{\textbf{DocReRank}-Base}  & \underline{\textbf{74.9}} & \textbf{64.0} & \textbf{67.2} & \textbf{74.8} & \underline{\textbf{70.8}} & \textbf{58.6} & \textbf{69.1} & 66.5 & \textbf{68.2} \\

\textit{\textbf{DocReRank}-Full}  & \textbf{74.3} & \underline{\textbf{67.0}} & \underline{\textbf{69.5}} & \underline{\textbf{76.7}} & \textbf{70.7} & \underline{\textbf{62.6}} & \underline{\textbf{69.5}} & \underline{\textbf{67.3}} & \underline{\textbf{69.7}} \\

\bottomrule
\end{tabular}
\caption{
\textbf{Performance on the ViDoReV2
Benchmark NDCG@10}  
}
\label{Table:vidoreV2_ncdg10}
\vspace{-0.4cm}
\end{table*}

\clearpage
\centering
\begin{table*}[ht!]
\centering
\footnotesize
\renewcommand{\arraystretch}{1.5} % Adjust row height
\setlength{\dashlinedash}{0.5pt}
\setlength{\dashlinegap}{0.5pt}
\setlength\tabcolsep{2.5pt} % Adjust column spacing
\begin{tabular*}{\textwidth}{@{\extracolsep{\fill}}lccccc} 
\toprule
\textbf{Benchmark} & \textbf{FinReport} & \textbf{FinSlides} & \textbf{TechReport} & \textbf{TechSlides} & \textbf{Avg} \\
\midrule
\textit{ColPali}                  & 39.6                     & 46.9                     & 67.9                     & 81.7                     & 59.0                     \\

\textit{Qwen-VLM}                & 45.5                     & 62.1                     & 58.2                     & 63.3                     & 57.3                     \\

\textit{MonoQwen}                & \textbf{60.8}            & 67.8                     & 61.1                     & 83.4                     & 68.3                     \\

\textit{\textbf{DocReRank}-Base} & 58.6                     & \textbf{77.5}            & \textbf{80.9}            & \textbf{89.1}            & \textbf{76.5}            \\

\textit{\textbf{DocReRank}-Full} & \underline{\textbf{61.4}} & \underline{\textbf{77.6}} & \underline{\textbf{81.2}} & \underline{\textbf{89.4}} & \underline{\textbf{77.4}} \\

\addlinespace

\textit{ColQwen}                  & 44.7                     & 43.2                     & 73.4                     & 84.1                     & 61.4                     \\

\textit{Qwen-VLM}                & 47.7                     & 58.8                     & 57.6                     & 63.6                     & 56.9                     \\

\textit{MonoQwen}                & 59.0                     & 63.3                     & 60.7                     & 83.7                     & 66.7                     \\

\textit{\textbf{DocReRank}-Base} & \textbf{60.5}            & \underline{\textbf{73.7}} & \textbf{80.6}            & \textbf{89.5}            & \textbf{76.1}            \\

\textit{\textbf{DocReRank}-Full} & \underline{\textbf{64.8}} & \textbf{73.1}            & \underline{\textbf{81.3}} & \underline{\textbf{89.6}} & \underline{\textbf{77.2}} \\

\bottomrule
\end{tabular*}
\caption{
\textbf{Performance on the Real-MM-RAG
Benchmark recall@1}
}
\label{Table:real-mm-rag_recall1}
\vspace{-0.4cm}
\end{table*}

\vspace{1em}

\begin{table*}[ht!]
\centering
\footnotesize
\renewcommand{\arraystretch}{1.5} % Adjust row height
\setlength{\dashlinedash}{0.5pt}
\setlength{\dashlinegap}{0.5pt}
\setlength\tabcolsep{2.5pt} % Adjust column spacing
\begin{tabular*}{\textwidth}{@{\extracolsep{\fill}}lccccc} 
\toprule
\textbf{Benchmark} & \textbf{FinReport} & \textbf{FinSlides} & \textbf{TechReport} & \textbf{TechSlides} & \textbf{Avg} \\
\midrule
\textit{ColPali}                  & 64.7                     & 76.0                     & 90.1                     & 94.9                     & 81.4                     \\

\textit{Qwen-VLM}                & 77.1                     & 88.5                     & 86.8                     & 91.5                     & 86.0                     \\

\textit{MonoQwen}                & \underline{\textbf{82.3}} & 92.6                     & 93.8                     & \textbf{97.7}            & 91.6                     \\

\textit{\textbf{DocReRank}-Base} & 81.6                     & \textbf{92.7}            & \underline{\textbf{96.1}} & 97.5                     & \textbf{92.0}            \\

\textit{\textbf{DocReRank}-Full} & \textbf{82.1}            & \underline{\textbf{93.1}} & \textbf{95.5}            & \underline{\textbf{97.8}} & \underline{\textbf{92.1}} \\

\addlinespace

\textit{ColQwen}                  & 74.9                      & 71.5                      & 93.2                      & 96.5                      & 84.0                      \\

\textit{Qwen-VLM}                & 80.9                      & 82.5                      & 86.6                      & 91.7                      & 85.4                      \\

\textit{MonoQwen}                & 89.0                      & 85.7                      & 94.3                      & 98.1                      & 91.8                      \\

\textit{\textbf{DocReRank}-Base} & \textbf{89.4}             & \textbf{85.9}             & \underline{\textbf{97.2}} & \textbf{98.2}             & \textbf{92.7}             \\

\textit{\textbf{DocReRank}-Full} & \underline{\textbf{90.0}} & \underline{\textbf{86.5}} & \textbf{97.1}             & \underline{\textbf{98.4}} & \underline{\textbf{93.0}} \\

\bottomrule
\end{tabular*}
\caption{
\textbf{Performance on the Real-MM-RAG
Benchmark recall@5}
}
\label{Table:real-mm-rag_recall5}
\vspace{-0.4cm}
\end{table*}

\vspace{1em}

\begin{table*}[ht!]
\centering
\footnotesize
\renewcommand{\arraystretch}{1.5} % Adjust row height
\setlength{\dashlinedash}{0.5pt}
\setlength{\dashlinegap}{0.5pt}
\setlength\tabcolsep{2.5pt} % Adjust column spacing
\begin{tabular*}{\textwidth}{@{\extracolsep{\fill}}lccccc} 
\toprule
\textbf{Benchmark} & \textbf{FinReport} & \textbf{FinSlides} & \textbf{TechReport} & \textbf{TechSlides} & \textbf{Avg} \\
\midrule
\textit{ColPali}                & 56.0                     & 66.1                     & 81.8                     & 90.0                     & 73.5                     \\

\textit{Qwen-VLM}               & 64.5                     & 78.2                     & 76.1                     & 80.5                     & 74.8                     \\

\textit{MonoQwen}               & \textbf{73.7}            & 82.3                     & 80.3                     & 92.1                     & 82.1                     \\

\textit{\textbf{DocReRank}-Base}& 72.5                     & \textbf{86.5}            & \textbf{89.9}            & \textbf{94.4}            & \textbf{85.8}            \\

\textit{\textbf{DocReRank}-Full}& \underline{\textbf{73.8}}& \underline{\textbf{86.9}}& \underline{\textbf{90.0}}& \underline{\textbf{94.5}}& \underline{\textbf{86.3}}\\

\addlinespace

\textit{ColQwen}                    & 64.6              & 61.6              & 85.6              & 91.9              & 75.9              \\

\textit{Qwen-VLM}                   & 68.6              & 72.9              & 76.0              & 80.7              & 74.6              \\

\textit{MonoQwen}                   & 76.7              & 76.5              & 80.6              & 92.5              & 81.6              \\

\textit{\textbf{DocReRank}-Base}    & \textbf{77.8}     & \textbf{80.8}     & \textbf{90.4}     & \textbf{94.9}     & \textbf{86.0}     \\

\textit{\textbf{DocReRank}-Full}    & \underline{\textbf{79.7}} & \underline{\textbf{80.9}} & \underline{\textbf{90.8}} & \underline{\textbf{95.0}} & \underline{\textbf{86.6}} \\

\bottomrule
\end{tabular*}
\caption{
\textbf{Performance on the Real-MM-RAG
Benchmark NDCG@10 }
}
\label{Table:real-mm-rag_ndcg10}
\vspace{-0.4cm}
\end{table*}

\clearpage
\vspace{1em}

\begin{table*}[ht!]
\footnotesize
\renewcommand{\arraystretch}{1.5} % Adjust row height
\setlength{\dashlinedash}{0.5pt}
\setlength{\dashlinegap}{0.5pt}
\setlength\tabcolsep{2pt} % Adjust column spacing
\hspace{-0.15cm}
\begin{tabular}{@{\extracolsep{\fill}}lccccccccc} 
\toprule
\textbf{Benchmark} & \textbf{Axa} & \textbf{Economics} & \textbf{Restaurant-rse} & \textbf{Restaurant-esg}  & \textbf{Biomedical} & \textbf{Economics-ML} & \textbf{Restaurant-ML} & \textbf{Biomedical-ML} & \textbf{Avg}\\
\midrule

\textit{ColPali}   & 55.0 & 53.4 & 51.5 & 55.2 & 57.8 & 47.6 & 52.5 & 55.6 & 53.6 \\

\hspace{0.1cm} \textit{Qwen-VLM}   & 61.8 & 47.9 & 57.7 & 60.8 & 56.4 & 49.8 & 58.0 & 56.1 & 56.1 \\

\hspace{0.1cm} \textit{FT on Col-HNDoc}   & 66.4 & 61.4 & 59.4 & 63.3 & 67.1 & 55.1 & 59.2 & 62.6 & 61.8 \\

\hspace{0.1cm} \textit{\textbf{DocReRank}-Base}   & \underline{\textbf{71.8}} & 64.7 & 58.3 & 68.8 & 68.0 & 58.3 & 61.2 & 62.9 & 64.2 \\

\hspace{0.1cm} \textit{\textbf{DocReRank}-B w Fin}   & 68.5 & 64.2 & \textbf{62.2} & 68.2 & \textbf{68.2} & \textbf{58.3} & \textbf{62.0} & 63.5 & 64.4 \\

\hspace{0.1cm} \textit{\textbf{DocReRank}-B w Fin\&Reph}   & 68.9 & \textbf{66.4} & 61.6 & \textbf{69.4} & \underline{\textbf{68.4}} & 60.1 & 62.5 & \underline{\textbf{64.7}} & \textbf{65.3} \\

\hspace{0.1cm} \textit{\textbf{DocReRank}-Full}   & \textbf{70.7} & \underline{\textbf{67.5}} & \underline{\textbf{63.4}} & \underline{\textbf{71.8}} & 66.8 & \underline{\textbf{62.8}} & \underline{\textbf{63.3}} & \textbf{63.7} & \underline{\textbf{66.2}} \\

\bottomrule
\end{tabular}
\caption{
\textbf{Dataset Ablation Study on ViDoReV2 Benchmark with ColPali}  
}
\label{Table:ablation_colpali}
\vspace{-0.4cm}
\end{table*}

\begin{figure*}[h!]
    \centering
    \includegraphics[width=1.0\textwidth]{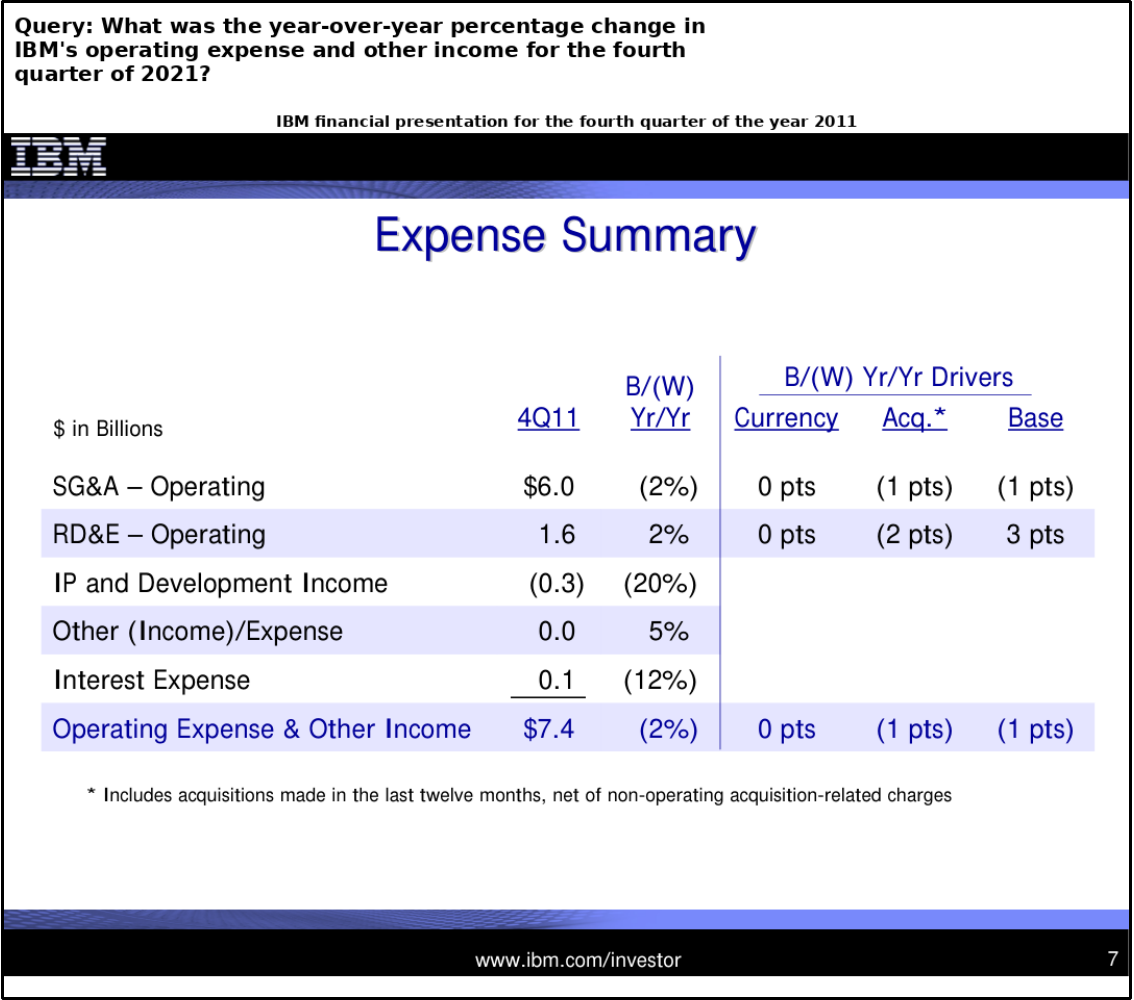} 
    \caption{\textbf{Reranker Failure Case (Wrong Year):} We show an example where the reranker ranked a page first, but it was labeled as negative (i.e., it does not answer the query). This case from FinSlides demonstrates that the model correctly identified relevant cues—such as "expenses," "operating," "year-over-year," and "Q4"—but failed on the year: the query asked about 2021, while the retrieved page was from 2011.}
    \vspace{-0.2cm}
    \label{fig_sup:fail_cases_fin_1}
\end{figure*}

\begin{figure*}[h!]
    \centering
    \includegraphics[width=1.0\textwidth]{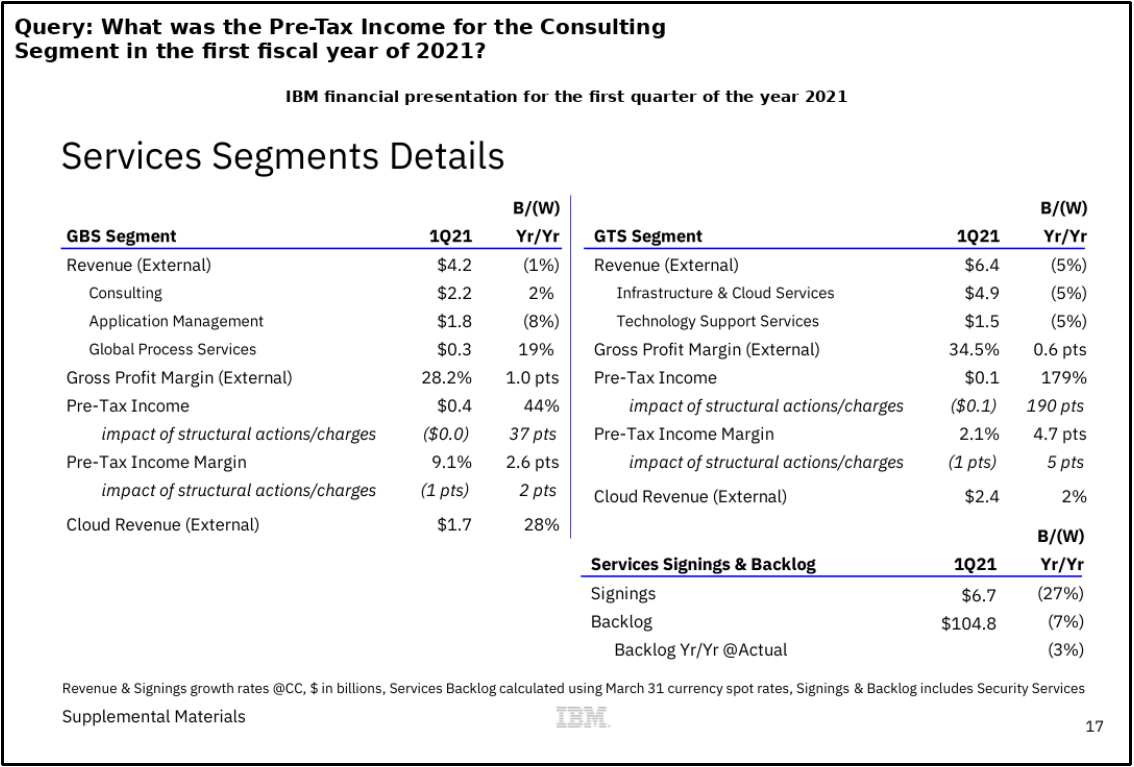} 
    \caption{\textbf{Reranker Failure Case (Wrong Business Segment):} We show an example where the reranker ranked a page first, but it was labeled as negative (i.e., it does not answer the query). This case from FinSlides demonstrates that the model correctly identified relevant cues—such as "first quarter 2021," "year-over-year," and "pre-tax income"—but failed on the business segment: the query asked about the consulting segment, while the retrieved page referred to the services segment.}
    \vspace{-0.2cm}
    \label{fig_sup:fail_cases_fin_2}
\end{figure*}

\end{document}